\documentclass[12pt,letterpaper]{article}                           

\usepackage{amsmath}
\usepackage{amssymb}

\usepackage{ifpdf}
\ifpdf
  \usepackage[pdftex, 
    pdftitle={On Abelian Gauge Symmetries and Proton Decay in Global F-theory GUTs}, 
    pdfauthor={Thomas W. Grimm and Timo Weigand}, 
    pdfsubject={String theory model building},
    bookmarksopen, bookmarksnumbered, bookmarksopenlevel=2]{hyperref}
\fi

\def\hybrid{\topmargin -20pt    \oddsidemargin 0pt
        \headheight 0pt \headsep 0pt
        \textwidth 6.25in       
        \textheight 9 in       
        \marginparwidth .875in
        \parskip 5pt plus 1pt 
          \jot = 1.5ex
   }
\hybrid
\numberwithin{equation}{section}
\numberwithin{table}{section}

\newcommand{\beq}{\begin{equation}}  \newcommand{\eeq}{\end{equation}}
\newcommand{\bal}{\begin{aligned}}   \newcommand{\eal}{\end{aligned}}
\newcommand{\bea}{\begin{eqnarray}}  \newcommand{\eea}{\end{eqnarray}}

\def\ov{\overline}

\newcommand{\bbP}{\mathbb{P}}

\newcommand{\nn}{\nonumber}

\newcommand{\fb}{\mathfrak{b}}
\newcommand{\fc}{\mathfrak{c}}

\newcommand{\cC}{\mathcal{C}}

\newcommand{\cL}{\mathcal{L}}

\newcommand{\cN}{\mathcal{N}}

\begin{document}

\baselineskip=14pt
\parskip 5pt plus 1pt

\vspace*{-1.5cm}
\begin{flushright}    
  {\small
  HD-THEP-11/10\\
  NSF-KITP-10-072\\
  }
\end{flushright}

\vspace{2cm}
\begin{center}        
  {\Large
  On Abelian Gauge Symmetries and Proton Decay \\[.2cm] in Global F-theory GUTs
  }
\end{center}

\vspace{0.75cm}
\begin{center}        
 Thomas W.~Grimm$^{a,c}\footnote{grimm@th.physik.uni-bonn.de}$  and Timo Weigand$^{b,c}$\footnote{t.weigand@thphys.uni-heidelberg.de}
\end{center}

\vspace{0.15cm}
\begin{center}        
  \emph{$^{a}$ Bethe Center for Theoretical Physics, \\ 
               Nussallee 12, 53115 Bonn, Germany } 
\\[0.15cm]
  \emph{$^{b}$ Institute for Theoretical Physics, University of Heidelberg,  \\ 
               Philosophenweg 19, 69120 Heidelberg, Germany } 
\\[0.15cm]
 \emph{$^{c}$ Kavli Institute for Theoretical Physics,  \\ 
                       Santa Barbara, CA 93106, USA } 
\\[0.15cm]

\end{center}

\vspace{1.0cm}


\begin{abstract}

The existence of abelian gauge symmetries in four-dimensional F-theory
compactifications depends on the global geometry of the internal Calabi-Yau 
fourfold and has important phenomenological consequences.
We study conceptual and phenomenological aspects of such $U(1)$ symmetries
along the Coulomb and the Higgs branch. As one application we examine
abelian gauge factors arising after a certain global restriction of
the Tate model
that goes beyond a local spectral cover analysis. In $SU(5)$ GUT models this
mechanism enforces a global $U(1)_X$ symmetry that prevents
dimension-4 proton decay
and allows for an identification of candidate right-handed
neutrinos. We invoke
a detailed account of the singularities of Calabi-Yau fourfolds and
their mirror duals starting from
an underlying $E_8$ and $E_7 \times U(1)$ enhanced Tate model.
The global resolutions and deformations of these singularities can be
used as the appropriate framework to analyse F-theory GUT models.

\end{abstract}

\clearpage



\newpage

\tableofcontents

\section{Introduction}

The prospects of F-theory for the construction of realistic GUT models \cite{Donagi:2008ca,Beasley:2008dc,Beasley:2008kw,Donagi:2009ra} 
have recently revived interest also in more formal aspects of F-theory compactifications on Calabi-Yau fourfolds.
While many phenomenological challenges of GUT model building  can be and have been addressed already at the level 
of local models\footnote{For an incomplete list of references see e.g. \cite{Heckman:2008qa}-\cite{King:2010mq}.}
, important issues remain which defy a treatment without reference to the global 
properties of the compactification. Among these are most notably almost all questions pertinent to abelian gauge symmetries.
Already the GUT breaking mechanism with the help of hypercharge flux, which is one the characteristics of the F-theory GUT 
models advanced in \cite{Donagi:2008ca,Beasley:2008dc,Beasley:2008kw,Donagi:2009ra}, is sensitive to global compactification 
data because such flux can only be turned on along two-cycles that are trivial in the homology of the full Calabi-Yau four-fold.
More fundamentally, the very definition of abelian gauge symmetries hinges upon global information. This phenomenon is well-known 
already in the context of perturbative Type II or heterotic string vacua, where St\"uckelberg couplings to axionic fields can 
degrade gauge symmetries to merely global selection rules below the mass scale of the gauge boson.

The global data of an elliptic Calabi-Yau fourfold is in general encoded in the Weierstrass model. In this paper we focus on elliptic fibrations that can be written globally in Tate form.
The localised gauge degrees of freedom can be read off from the singularities of the elliptic fiber  as reviewed in \ref{Tateform}. 
In constructing global examples it is important to have a method to explicitly resolve the singularities. This is crucial not only to control the topology of the compactification and to reliably compute the Euler characteristic, but also to determine the physical spectrum such as the number of $U(1)$ bosons below the Kaluza-Klein scale.
Such an explicit construction of a class of singular Calabi-Yau fourfolds and their resolution has been achieved in ref.~\cite{Blumenhagen:2009yv,Grimm:2009yu} 
in terms of toric 
geometry, extending the program of global toric Type IIB orientifold GUT model building advanced in \cite{Blumenhagen:2008zz}. 
See \cite{Marsano:2009ym,Marsano:2009gv,Marsano:2009wr} for F-theory models on a different singular Calabi-Yau fourfold. 
Very recently, a similar approach as in ~\cite{Blumenhagen:2009yv,Grimm:2009yu} to the explicit construction and resolution  of  Calabi-Yau fourfolds has been taken in \cite{Chen:2010ts}. The manifolds of \cite{Marsano:2009ym} and \cite{Blumenhagen:2009yv} appear also in the models \cite{Chen:2010tp}.

What has made the construction of three-generation $SU(5)$ GUT  ~\cite{Blumenhagen:2009yv,Grimm:2009yu,Marsano:2009ym,Marsano:2009gv,Marsano:2009wr}  and flipped $SU(5)$ GUT  \cite{Chen:2010ts,Chen:2010tp} models possible is in addition the 
description of gauge flux with the help of the spectral cover construction  \cite{Donagi:2008ca,Hayashi:2008ba,Donagi:2009ra,Hayashi:2009ge}. 
The latter can be thought of as the restriction of the Tate model to the vicinity of the $SU(5)$ GUT brane and 
as such necessarily discards some of the global information of the model. 
To decide in concrete global examples whether the spectral cover nonetheless captures the main aspects of the geometry correctly requires further tests.
For the examples of ~\cite{Blumenhagen:2009yv,Grimm:2009yu} a spectral cover based formula for the Euler characteristic has been compared to the value computed independently via toric geometry and match was found. This was taken as an indication that the spectral cover methods are applicable in these cases. Let us stress, however, that the spectral cover formula of ~\cite{Blumenhagen:2009yv} was not meant as an unambiguous method to compute the Euler characteristic in examples where independent computations are not available, but rather as a check on the applicability of the spectral cover methods to these Calabi-Yau manifolds. An even stronger indicator will be given in section \ref{GlobalSCC} of the present article via mirror symmetry, which in suitable  examples  exchanges the gauge group and spectral cover group.

As a special class of constructions, the so-called split spectral cover was 
discussed in \cite{Marsano:2009ym,Marsano:2009gv}  
(and applied therein and in \cite{Blumenhagen:2009yv,Grimm:2009yu}) 
as a method to implement abelian selection rules in the gauge theory 
of the GUT theory, e.g. $U(1)_X$ in the decomposition $SO(10) \rightarrow SU(5) \times U(1)_X$. 
These can forbid unwelcome couplings such as the dangerous dimension-4 proton decay operators.
Given the global nature of abelian gauge symmetries, however, it has continued to be an 
open question whether these selection rules are unbroken also in the full global model. 
Besides, the nature of $SU(5)$ gauge singlets playing the role of right-handed neutrinos 
has remained elusive in the spectral cover picture because these states are not localised 
on the GUT brane and are thus beyond the actual scope of the spectral cover.

In fact, the result of the analysis put forward in the present paper is that questions 
such as the existence of (massive) $U(1)$ gauge symmetries cannot be answered in a 
satisfactory manner from a spectral cover perspective. Rather, we find a method to ensure 
the presence of abelian symmetries and thus to remove dangerous dimension-4 proton decay 
operators by considering special limits of the Tate model of the compact Calabi-Yau fourfold itself. We call the resulting construction $U(1)$-\emph{restricted Tate models}.
Note that quite recently, it has been argued in  \cite{Hayashi:2010zp} from the perspective 
of the monodromies of the F-theory model that $U(1)_X$ is generically broken in split spectral 
cover models. Our approach has been independent of these findings and offers a complimentary 
view on the breaking of $U(1)_X$. We also describe a way to ensure that  $U(1)_X$ is preserved as a  - possibly massive  - symmetry below the Kaluza-Klein scale in global models.
Implementing this construction  in concrete examples is of phenomenological importance not only in order to guarantee the existence of abelian selection rules advertised in the split spectral cover context; rather, the very computation of individual chiral matter indices in split spectral cover constructions can be invalidated if the $U(1) $ symmetries are higgsed at a global level. Determining whether this is the case requires a global analysis of the type put forward in the sequel.

In section \ref{fiber-restriction} we start from a Tate model with gauge group enhanced to 
$E_8$ along the divisor $S$. The final model  with gauge group $G \subset E_8$ appears 
as a deformation thereof. In section \ref{E8-Cartan} we study the Cartan $U(1)$s within 
$E_8$ by moving to the Coulomb branch via direct resolution and identifying the extended 
Dynkin diagram within the resolved fiber. We argue that the deformation of the Tate model 
to gauge group $G$ generically higgses all $U(1)$ symmetries within $H= E_8/G$. 
In non-generic cases, however, some $U(1)$ symmetries remain. 
The gauge flux associated with the Cartan generators of $H$ are most conveniently studied along the Coulomb branch. After the deformation
they describe data of a truly non-abelian $H$-bundle.
A special role is played by 
the extended node in the extended Dynkin diagram. In section \ref{Massive} we argue that 
this node carries information about massive $U(1)$ gauge symmetries that have acquired a 
Kaluza-Klein scale mass via the F-theoretic analogue of the Type II St\"uckelberg mechanism. 
This clarifies the F-theory fate of the ubiquitous Type II $U(1)$s.

We then specialise to the $U(1)_X$ symmetry in $SU(5)$ GUT models. 
In section \ref{SplitSCC} we recall its local description in terms of  
an $S[U(4) \times U(1)_X]$ spectral cover. We argue that generically this symmetry is 
higgsed by VEVs of $U(1)_X$ charged $SU(5)$ GUT singlets localised away from the GUT 
brane and invisible to the spectral cover analysis. 
In this case effective dimension-4 proton violating couplings may ruin the model 
in a way inaccessible from the spectral cover perspective.
We then resolve the problem of dimension-4 proton decay in section \ref{Globalfac} by 
promoting the split of the spectral cover to a global restriction of the sections 
appearing in the Tate model.\footnote{Note that 
six-dimensional F-theory compactifications with a restricted Tate model and additional $U(1)$ factors have 
been studied from the perspective of heterotic/F-theory duality in refs.~\cite{Aldazabal:1996du,Candelas:1997pv,Berglund:1998va}.} This gives an unambiguous way to determine the presence 
of $U(1)_X$ by detection of a curve of $SU(2)$ enhancement on the $I_1$ divisor 
of the discriminant. We identify this curve as the proper localisation curve of 
the states charged only under $U(1)_X$ which have the correct quantum numbers to 
play the role of right-handed neutrinos. This resolves also the puzzle of 
the localisation of neutrinos in the spectral cover context \cite{Blumenhagen:2009yv,Marsano:2009gv}. 
The appearance of an extra abelian gauge symmetry is further 
confirmed in section \ref{Bosons}, where we resolve the singular curve. 
 In the spirit of M/F-duality the resulting increase in $h^{1,1}$ indicates a $U(1)$ boson which is massless in absence of gauge flux. The latter can render the abelian symmetry massive, which then survives only as a global symmetry valid below the Kaluza-Klein scale. For applications such as the engineering of $U(1)_X$ this is exactly what one is interested in for model building.  
We furthermore show that the underlying gauge symmetry in the Tate model is $E_7 \times U(1)$, as opposed to the previously described $E_8$ for the generic Tate model.
A drawback of the $U(1)$ restriction, however, is seen to be a significant decrease in the Euler characteristic of the fourfold.
This challenges the construction of models satisfying the D3-tadpole constraint.
Finally, in section \ref{Orientifolds} we describe how the appearance of $U(1)_X$  can be 
further understood in analogy with Type IIB orientifolds, where the restricted Tate model 
describes brane-image brane splitting.

In section \ref{GlobalSCC} we study in greater detail the deformation of the 
underlying $E_8$ Tate model to a compactification with $G \subset E_8$, thereby putting 
the logic of the local spectral cover approach into perspective with global models. 
This section fills in the technical details for some of the claims in section \ref{fiber-restriction} 
and stresses the use of mirror symmetry to study the deformations.
In section \ref{SCCs} we work out the natural appearance of the underlying $E_8$ structure 
for a certain class of F-theory models based on a ${\mathbb P}_{1,2,3}[6]$ Tate model. 
In fact the breaking of $E_8$ to $G$ along a divisor via a bundle with structure group $H=E_8/G$ 
is recovered entirely in terms of the Tate model. These observations are independent of a local 
gauge theory perspective or a heterotic dual. 
This picture is corroborated further in section \ref{mirror_spec} with the help of an analysis 
of the mirror dual Calabi-Yau fourfolds and their gauge enhancements. It is found that in the 
mirror dual precisely those gauge groups appear which are responsible for the unfolding 
of $E_8$ to the various codimension loci of singularity enhancement.
Finally in section \ref{GlobalSCCs} this logic is applied to the mirror dual of the 
restricted Tate model corresponding to a split spectral cover based on an underlying $E_7 \times U(1)$.

\section{Compact Calabi-Yau fourfolds and abelian gauge symmetries} \label{fiber-restriction}

\subsection{Complete-intersecting fourfolds and the Tate form} \label{Tateform}

To set the stage we introduce the Calabi-Yau fourfolds on which we compactify 
F-theory. The class of fourfolds we consider is rather general so as to cover 
the geometries which have been recently used in the study of 
F-theory GUT models in refs.~\cite{Blumenhagen:2009yv,Grimm:2009yu,Chen:2010ts}. 
We explicitly realise the Calabi-Yau fourfold $Y$ 
via \textit{two} hypersurface constraints 
\beq \label{two_constr}
   P_{\rm base}(y_i) = 0 ,\qquad P_{\rm T}(x,y,z;y_i)=0 
\eeq
in a six-dimensional projective or toric ambient space. Here $P_{\rm base}$
is the constraint of the base $B$ which is independent of the coordinates 
$(x,y,z)$ of the elliptic fiber. This more general setting 
also includes hypersurfaces encoded by a single constraint $P_{\rm T} =0$ if 
$P_{\rm base}$ is chosen to be trivial.  
$P_{\rm T}=0$ is the constraint that describes the structure of the elliptic fibration. We consider fibrations that can be given in 
Tate form,
\bea \label{Tate1}
    P_{\rm T} = x^3 - y^2 + x\, y\,  z\, a_1 + x^2\, z^2\, a_2 + y\, z^3\,a_3   + x\, z^4\, a_4  +  \, z^6\, a_6\ = 0,
\eea
where $(x,y,z)$ are coordinates of the torus fiber. In the sequel we will often 
be working with the inhomogeneous Tate form by setting $z=1$. The $a_n(y_i)$  
depend on the complex coordinates $y_i$ of the base $B$ such as to form
sections of $K_B^{-n}$, with $K_B$ being the canonical bundle of the base $B$.  
Setting all $a_n=1$ one finds that \eqref{Tate1} reduces to the elliptic fiber $\bbP_{1,2,3}[6]$. 
A fibration based on  a representation of the elliptic curve as $\bbP_{1,2,3}[6]$ is called $E_8$ 
fibration for reasons that will be become clearer in section \ref{SCCs}.

To put the ansatz (\ref{Tate1}) into perspective we recall that most generally every elliptic fourfold with section admits a description as a Weierstrass model
\bea
\label{Weierstrass}
P_{\rm W } = x^3 - y^2 + f x z^4 + g z^6 = 0.
\eea
Clearly, every Tate model (\ref{Tate1}) can be brought into this form via the relation
\beq
\label{fg}
 f=-\frac{1}{48}( \beta_2^2 -24\, \beta_4), \qquad \quad g=-\frac{1}{864}( -\beta_2^3 + 36 \beta_2 \beta_4 -216 \, \beta_6) ,
\eeq 
where
\bea
\label{beta}
  \beta_2 = a_1^2 + 4 a_2 ,\quad 
  \beta_4 = a_1 a_3 + 2\, a_4 ,\quad
  \beta_6 = a_3^2 + 4 a_6  .
\eea
 In turn the Tate model is a specialisation of the Weierstrass model (\ref{Weierstrass}), 
which emerges naturally in toric elliptic fibrations.
 We will focus on this class of elliptic fibrations.

The sections $a_n$ encode the discriminant $\Delta$ of the elliptic fibration given by
\beq
  \Delta = -\tfrac14 \beta_2^2 (\beta_2 \beta_6 - \beta_4^2) - 8 \beta_4^3 - 27 \beta_6^2 + 9 \beta_2 \beta_4 \beta_6.
\eeq
The discriminant locus $\Delta$ 
may factorise with each factor describing the location of a 7-brane on a divisor 
$D_k$ in $B$. 
The precise gauge group along $D_k$ is encoded in the 
vanishing degree $\delta(D_k)$ of $\Delta$ and the vanishing degrees $\kappa_n(D_k)$ of the $a_n$,
\bea \label{TateSUn}
&&  
   a_1 = \mathfrak{b}_5 w^{\kappa_1} , \quad
   a_2 = \mathfrak{b}_4 w^{\kappa_2} , \quad
   a_3 = \mathfrak{b}_3 w^{\kappa_3} , \quad \nn \\
  &&  a_4 =  \mathfrak{b}_2 w^{\kappa_4} , \quad
   a_6 =  \mathfrak{b}_0 w^{\kappa_6} \ , \quad \Delta = \Delta' w^\delta\ ,
  \eea
as classified in Table~2 of ref~\cite{Bershadsky:1996nh}.
For example, for an $SU(5)$ gauge group along the divisor $S: w=0$, where $w$ is one of the base 
coordinates $y_i$, the $\kappa_n$ are given by
\beq
(\kappa_1, \kappa_2, \kappa_3, \kappa_4,\kappa_6) = (0,1,2,3,5).
\eeq
The sections $\mathfrak{b}_n$ generically depend on all coordinates $(y_i,w)$ 
of the base $B$ but do not contain an overall factor of $w$. Note that 
in an $SU(5)$ GUT model the $\mathfrak{b}_n$ identify the so-called matter 
curves along which zero modes charged under the $SU(5)$ gauge group are localised. 
The matter curves for the ${\bf 10}$  and  ${\bf 5}$ representations,
\bea \label{curve10}
&&    P_{\bf 10} : \quad z=w=0 \quad \cap  \quad \mathfrak{b}_5 = 0, \\
&&    P_{\bf 5} : \quad z=w=0 \quad \cap \quad  \mathfrak{b}_3^2 \mathfrak{b}_4 - \mathfrak{b}_2 \mathfrak{b}_3 \mathfrak{b}_5 + \mathfrak{b}_0 \mathfrak{b}_5^2 = 0, \nn
\eea
are the curves of singularity enhancement to $SO(10)$ and, respectively, $SU(6)$.
Note that for generic 
choice of sections $\mathfrak{b}_i$ the $ P_{\bf 5}$-curve does not factorise so that all matter 
in the fundamental representation, both ${\bf \ov 5}_M$ and the Higgs ${\bf 5}_H + {\bf \ov 5}_H$ 
are localised on the same curve. It will be crucial task of section \ref{U(1)-restricted} to identify a class of 
compact Calabi-Yau fourfolds for which the $ P_{\bf 5}$-curve splits into two curves hosting 
${\bf \ov 5}_M$ and ${\bf 5}_H + {\bf \ov 5}_H$, respectively.

It is important to stress 
that in the case of such a higher degeneration, not only the elliptic fibration will be 
singular, but rather the Calabi-Yau fourfold itself.  We will call this singular 
fourfold $Y_G$. The singularities of gauge group $G$ can be resolved into a non-singular Calabi-Yau fourfold $\ov Y_G$. This 
has been demonstrated explicitly for the $SU(5)$ GUT examples in refs.~\cite{Blumenhagen:2009yv,Grimm:2009yu}. 
The existence of such a resolved space is crucial to determine  the topological data, as e.g.~required
for tadpole cancellation. $\ov Y_G$ is also plays a vital role in 
the study of Cartan symmetries and hence will be discussed more thoroughly
in the next section.

\subsection{Cartan $U(1)$s and underlying $E_8$ structure}
\label{E8-Cartan}

In this section we study the connection between the Cartan $U(1)$ potentials and fluxes for 
the gauge theory on the brane $S$ and the geometry of $Y_G$. Note that the globally varying dilaton profile
makes it hard to obtain the spectrum and effective action of an 
F-theory compactification on a singular $Y_G$. What rescues us is the duality of 
F-theory to an M-theory reduction, which leads to an identification 
of $U(1)$ symmetries in geometric terms. Our strategy is to 
use the fact that in the M-theory reduction one can access the 
Coulomb branch of the gauge theory in which $G$ is broken to $U(1)^{{\rm rk}(G)}$.
In fact this branch is attained upon resolving $Y_G$ into $\ov Y_G$ as follows:

The singular fiber of $Y_G$ over the discriminant $\Delta$
contains a tree of zero-volume $\mathbb P^1$s, 
which intersect as the nodes of the \emph{extended} Dynkin diagram of $G$ in the fiber of the fourfold. 
Resolution corresponds to blowing-up the $\mathbb P^1$s to non-zero volume. 
More precisely, one can resolve the 
generic $G$ singularity over $S$ by introducing ${\rm rk}(G)$ blow-up divisors $D_i$, $i =1, \ldots {\rm rk}(G)$, which 
are $\mathbb{P}^1$ bundles over $S$. The extended node of the affine Dynkin diagram is obtained as the linear combination 
\beq \label{extended_node}
D_0 = \hat S - \sum_i a_i D_i, 
\eeq
where the divisor $\hat S$ in $\ov Y_G$ is the elliptic 
fibration over $S$, and $a_i$ are the Dynkin numbers of the 
Dynkin node associated with $D_i$ (see, e.g.~\cite{Katz:1996th}). 
Explicit constructions of these $D_i$ are known for 
various compact Calabi-Yau manifolds \cite{Candelas:1996su,Bershadsky:1996nh,Candelas:1997eh}, including Calabi-Yau fourfolds 
relevant for GUT model building \cite{Blumenhagen:2009yv,Grimm:2009yu}. 
Let us denote by $\omega_i,i=0,...,{\rm rk}(G)$ the elements of $H^{2}(\ov Y_G,\mathbb{Z})$ which are Poincar\'e dual 
to $D_0$ and the blow-up divisors $D_i$. 
The intersections of the fiber $\mathbb{P}^1$s are captured by the 
identify 
\beq \label{Cartan-intersect}
    \int_{\ov Y_G} \omega_i \wedge \omega_j \wedge \tilde \omega = - C_{ij} \int_S \tilde \omega\ ,
\eeq
where $\tilde \omega$ is a four-form on the base $B$ and $C_{ij}$ is the extended Cartan matrix of $G$.
The appearance of $C_{ij}$ links the group theory of $G$ with the intersection theory on the 
resolved fourfold $\ov Y_{G}$. The two-forms $\omega_i$ are thus in direct relation with the 
simple roots of the gauge group $G$. Note that the integral \eqref{Cartan-intersect} 
localises onto $S$ in agreement with the assertion that all  
non-abelian gauge dynamics localises onto $S$.

The connection between the gauge theory and geometry can be exploited further by noting that
for a Tate model of the form (\ref{Tate1}) the gauge enhancements follow a structure 
inherited from an underlying $E_8$ symmetry into which $G$ can be embedded.
This $E_8$ structure emerges naturally in the context of the 
globally realised Tate models in projective or toric ambient spaces in which it 
is possible to enhance the gauge symmetry to $E_8$ along the divisor $S$ 
by a suitable choice of sections $a_i$. 
This leads to a singular fourfold $Y_{E_8}$. For the examples 
relevant to this work one can show explicitly that $Y_{E_8}$ can be resolved into
a Calabi-Yau manifold $\ov Y_{E_8}$.
The original fourfold $Y_G$ is a deformation of $Y_{E_8}$ such that along $S$ 
the underlying $E_8$ is broken to $G= E_8/H$, where $G$ is the commutant of $E_8$ 
a group $H$. Physically this corresponds to 
a recombination of the seven-branes. The details of this $E_8$ 
enhancement and the deformations to $G \subset E_8$ are described in section \ref{SCCs}.
We hasten to add, though, 
that more general examples of Weierstrass
models with higher rank gauge groups that cannot 
be embedded into $E_8$ are known (see e.g. the recent \cite{Kumar:2009ac,McOrist:2010jw}). 
It would be important to understand the generalisation of the $E_8$ construction 
to these more general situations.

In order to not directly work with singular geometries, one can also 
attempt to interpret the transition between the two resolved 
Calabi-Yau fourfolds $\ov Y_{E_8}$ and $\ov Y_{G}$. 
As for $Y_G$  one moves to the Coulomb 
branch of $E_8$ via resolution of $Y_{E_8}$ into  $\ov Y_{E_8}$ by introducing 
the resolution divisors $D^{E_8}_i$, $i=0,\ldots,8$.
On $\ov Y_{E_8}$ the group theory 
of $E_8$ is realised via global fourfold intersections as in \eqref{Cartan-intersect}, now for the dual 2-forms $\omega_i^{E_8}$.
One can divide the resolution divisors $D^{E_8}_i$ into two sets $D_i^G$ and  $D_i^H$ such that the 
dual 2-forms intersect as the respective Dynkin diagrams of the two commuting $E_8$ subgroups $G$ and $H$ in $E_8 \rightarrow G \times H$.
This makes the Cartan generators of the commutant $H$ of 
$G \subset E_8$ visible in terms of the two-forms $\omega^H_i$ with 
dual divisors $D_i^H$.

One  may think of the transition from $\ov Y_{E_8}$ to $\ov Y_{G}$ as follows: 
The deformation introduces monodromies\footnote{The role of monodromies in local 
F-theory models has been discussed in \cite{Hayashi:2009ge,Bouchard:2009bu,Marsano:2009gv,Hayashi:2010zp}.} for 
the individual $\mathbb{P}^1$ fibers of the $D_i^H$. 
Generically this higgses $H$ along $S$ completely. 
The $\mathbb{P}^1$ fibers of $D_i^H$ have been pulled off $S$ such 
as to lie in the fiber over the remaining $I_1$ locus which 
intersects $S$ along curves. 
Due to the monodromies these $\mathbb P^1$s are no longer independent, 
corresponding to the higgsing of the Cartan group $H$.
The change from $\ov Y_{E_8}$ to $\ov Y_{G}$ is thus captured by a geometric 
transition, in which the resolving K\"ahler volumes are replaced by 
complex structure deformations. Schematically one can summarise this as
\beq
\label{rec-tab}
    \begin{array}{lcl}Y_{E_8} & \xrightarrow{\ \ \text{brane recomb.}\ \ } &   Y_G \\
                      \downarrow &  & \downarrow \\
                  \ov Y_{E_8} & \xrightarrow{\ \ \text{geom.~transition}\ \ }  & \ov Y_G \end{array}
\eeq
In section \ref{mirror_spec} we will show that starting from a reference 
geometry $\ov Y_{E_8}$ the dual group $H$ and its splittings 
can be studied using mirror symmetry for $\ov Y_{G}$.   

Having collected some geometric properties of $\ov Y_{G}$
and $\ov Y_{E_8}$ we now turn to the discussion of 
Cartan $U(1)$s and gauge fluxes. 
In M-theory the gauge fields $A^i$ in the Cartan algebra 
of $G$ arise by reduction of $C_3$ along the two-forms $\omega^G_i, i=0,...,{\rm rk}(G)$.
Note that only ${\rm rk}(G)$ of these two-forms are cohomologically independent
so that the expansion reads  
\beq \label{C3exp}
C_3 =  \sum^{{\rm rk}(G)}_{i=1} A^i \wedge \omega^G_i + \ldots\ .
\eeq
In order to find the actual Cartan $U(1)$s one first has to
introduce the linear combinations of $\omega^G_i$ representing the Cartan 
generators of $G$.
Recall that the $\omega_i^G$ correspond to the simple roots and intersect as in \eqref{Cartan-intersect}.

For the most generic deformation to $\ov Y_G$, the Cartan $U(1)$s of $H$ are 
higgsed completely in the course of the recombination process
of (\ref{rec-tab}). This corresponds to the maximal possible monodromy group 
acting on the  $\mathbb P^1$ fibers of $D_i^H$. However, some of the Cartan 
elements of $H$ may survive as massless $U(1)$ gauge symmetries in the full 
compactification if the higgsing is incomplete. One of the results of our analysis 
is to identify in section \ref{Globalfac} such a non-generic deformation 
of $Y_{E_8}$ to $Y_G$ which does leave a certain $U(1)$ unhiggsed. 

Since it will become relevant in that context, we now recall how to identify 
 $U(1)$ factors  in a full Weierstrass model. This gives an 
unambiguous way to determine the total rank of the brane gauge groups.
It is sufficient to count the number of $U(1)$ factors along  the Coulomb branch $\ov Y_G$.
The existence of an abelian gauge group below the 
Kaluza-Klein scale then hinges upon the availability of harmonic 
two-forms on the resolved Calabi-Yau fourfold $\ov Y$ along which the M-theory 
three-form can be reduced.\footnote{In addition, the reduction along 
harmonic $(2,1)$-forms of the base $B$ of $Y$ 
leads to so-called bulk $U(1)$s which are the equivalent of 
the Ramond-Ramond $U(1)$s obtained in a Type IIB reduction of $C_4$.} 
Not all these two-forms will correspond to four-dimensional $U(1)$
gauge bosons. Firstly, there are $h^{1,1}(B)$ elements of $H^2(B,\mathbb{Z})$
which lead to chiral K\"ahler moduli of the F-theory compactification. Secondly, 
there is one element in $H^{2}(\ov Y_G,\mathbb{Z})$ which corresponds to 
the class of the elliptic fiber and captures the 
extra metric degrees of freedom in the lift of a three-dimensional
M-theory compactification to a four-dimensional F-theory compactification.
In summary, the number of abelian brane gauge symmetries in the Coulomb branch 
is now given by
\beq
\label{nv}
  n_v = h^{1,1}(\ov Y_G) - h^{1,1}(B) - 1. 
\eeq
Note that  if $Y_G$ gives rise to the non-abelian gauge symmetry $G$, 
then the number of extra $U(1)$ factors that are not Cartan elements 
of $G$ is given by $\tilde n_v = n_v - rk(G)$.

To end this section, let us note that the construction of $\ov Y_{E_8}$
allows us to think about the description of gauge flux as follows: 
On $\ov Y_{E_8}$ a  subclass of gauge flux derives from the 
Cartan generators of $E_8$. The gauge flux within that class
leaving $G$ intact is of the form
\beq
\label{Hflux}
 G_4 = F_2^{(i)} \wedge \omega_i^H + \ldots ,
\eeq
where $\omega_i^H$ are the 2-forms dual to the divisors $D_i^H$ introduced above.
This can be made concrete when considering for example an $SU(5)$ GUT model 
on $S$. To define $G_4$ on $Y_{E_8}$ one first introduces a basis of 
fundamental weights of $H = SU(5)_{\perp}$ as
\beq \label{fund_weights}
   \lambda_{i} = \sum_{k=1}^{i} \omega_{5-k} \ , \quad i = 1,\ldots 5 \ ,
\eeq  
where the $\omega_i$ correspond to the simple roots of $E_8$ with intersection 
\eqref{Cartan-intersect}, including the extended node $\omega_0$. For the 
definition of the flux $G_4$ it is convenient to also introduce a `dual' 
basis of two-forms $\lambda^*_i$ corresponding to the dual weights as linear combinations
of the $\omega_i$. One demands that these satisfy 
\beq \label{dual-weights}
   \int_{\ov Y_{E_8}} \lambda^j \wedge \lambda^*_i \wedge \tilde \omega = \delta_i^j \int_S \tilde \omega \ . 
\eeq
The four-form flux $G_4$ can then be defined as 
\beq
   G_4 = \sum_{i=1}^5 F^{(i)}_2 \wedge \lambda_i^* \ .  
\eeq

To determine the analogous flux for $Y_G$ one 
therefore has to trace back (\ref{Hflux}) under the geometric transition from $\ov Y_{E_8}$ 
to $\ov Y_{G}$. For the most generic deformation to $\ov Y_G$ which higgses $H$ completely, the formerly
abelian fluxes become data of a 
genuinely non-abelian $H$-bundle. A precise study of these transitions is 
beyond the scope of this work, and will be presented elsewhere \cite{progress1}. 
After the deformation the $G_4$ flux is no longer of the simple form (\ref{Hflux}) because 
it will in general not be a sum of four-forms that can be written as a product of two two-forms.  However, 
in the transition from $\ov Y_{E_8}$ to $\ov Y_{G}$ the number of 
four-forms in $H^{2,2}(\ov Y_G)$ increases significantly.
These new four-forms indeed cannot be represented as a wedge of 
two non-trivial two-forms on $\ov Y_{G}$. Fluxes of this type are known to appear in the 
superpotential rather then the D-term (see ref.~\cite{Grimm:2009ef} for a recent 
discussion). This is in agreement with the fact that after the deformation the Cartan 
$U(1)$s of  $H$ are higgsed so that no field-dependent Fayet-Iliopoulos term can arise 
from the fluxes. If on the other hand a certain $U(1) \subset H$ survives, there exists 
abelian gauge flux associated with that $U(1)$ on $Y_G$, and we will argue in 
section \ref{Bosons} for the presence of an associated Fayet-Iliopoulos term.

\subsection{Massive $U(1)$s and the extended node}
\label{Massive}

In the previous section we discussed the appearance of the 
Cartan $U(1)$s in the breaking of $G$ to $U(1)^{{\rm rk}(G)}$.
In the Kaluza-Klein expansion we have used ${\rm rk}(G)$ independent 
forms $\omega^G_i$. However, the local geometry sees an additional two-form associated with the extended node 
of the Dynkin diagram \eqref{extended_node}. This extra form does not extend to an element of $H^2(\ov Y_G, \mathbb Z)$. In the language of resolving 
$\mathbb{P}^1$ fibers one thus finds a homological relation 
among the nodes $e_i$ of the extended Dynkin diagram and 
the elliptic fiber $e$. Geometrically this means that there exists 
a three-chain $\cC$ with boundary $\partial \cC$ given by   
\beq \label{chain}
   \partial \cC = \hat e,\qquad   \hat e = e - \sum_{i=0}^{{\rm rk}(G)} a_i e_i ,
\eeq
where $a_0=1$ for the $\mathbb{P}^1$ $e_0$ associated with the extended node. 
One can now attempt to include an additional two-form $\hat \omega$ in the 
reduction of $C_3$ which is non-closed and precisely captures the
relation \eqref{chain} as
\beq \label{domega}
    d \hat \omega =  \Psi, \qquad \int_{\cC} \Psi = \int_{\hat e} \hat \omega = 1.
\eeq 
It is possible to systematically include such non-closed and exact 
forms in the dimensional reduction.\footnote{See refs.~\cite{Gurrieri:2002wz,Grana:2005ny,Benmachiche:2006df} for 
initial discussions of $\cN=2$ and $\cN=1$ non-Calabi-Yau reductions. The $U(1)$ sector 
of Type II compactifications on non-K\"ahler manifolds has also been studied in detail in ref.~\cite{Grimm:2008ed}. }
This leads to compactification on non-Calabi-Yau manifolds.
To understand this in our context we first consider the M-theory compactification on an 
elliptic fibration that features $\hat \omega$ and later take the F-theory limit of vanishing fiber.
In particular we also include $\hat \omega$ in the 
expansion of a globally defined two-form $J = \hat v \, \hat \omega +\ldots $. 
This implies that $dJ$ does not vanish and one would interpret this as an 
M-theory compactification on a non-K\"ahler manifold. 
Moreover, $\hat \omega$ and $\Psi$ also appears in the expansion of $C_3$ as 
\beq \label{C3_exp_c}
   C_3 = \hat A \wedge \hat \omega + c\, \Psi + \ldots, 
\eeq 
where $\hat A$ is a vector and $c$ is a scalar in the non-compact dimensions.
The condition \eqref{domega} directly leads to the appearance of the 
covariant derivative $\mathcal{D} c = dc + \hat A$ in the field strength $F_4$ of $C_3$ 
as
\beq \label{gaugedF4}
   F_4 = \mathcal{D} c \wedge \Psi + \hat F \wedge \hat \omega + \ldots 
\eeq 
This is the M/F-theory analogue of a St\"uckelberg term known in weakly coupled Type II theories and
 implies that the $U(1)$ field can be rendered massive 
by absorbing the scalar $c$ with a gauge transformation. 
Now we take the limit of vanishing fiber in the lift 
from M-theory to F-theory.  The non-K\"ahler-ness disappears in the 
limit $\hat v \,  \rightarrow 0 $ while the extra $U(1)$ fields remain as massive fields at the Kaluza-Klein scale. Hence, one expects that these 
massive $U(1)$ factors play a prominent role in an F-theory 
compactification. 

A clean interpretation of the geometrically 
massive $U(1)$s in \eqref{gaugedF4} is provided in 
F-theory compactifications with an $SU(N)$ gauge group 
which admit a well-defined orientifold limit. In the 
orientifold picture one actually starts from a D7-brane 
construction with a $U(N)=SU(N)\times U(1)$ gauge group.
Such examples arise if a stack of $N$ D7-branes is identified 
with an image stack in the Calabi-Yau threefold 
double cover $Z$ of $B$. If these two stacks wrap 
four-cycles in different homology classes, the $U(1)$ in 
the decomposition of $U(N)$ becomes massive since a scalar 
appearing in the expansion of the R-R two-form $C_2$ is gauged \cite{Jockers:2004yj}.
This can be inferred from the St\"uckelberg term of the D7-brane Chern-Simons action
\beq \label{CSact}
  S_{CS} \supset \int \hat F \wedge C_6 , 
\eeq
where $C_6$ it the R-R six-form dual to $C_2$, and $\hat F$ is the
field strength of the brane $U(1)$. The associated mass term is purely geometric 
and independent of any fluxes. Hence, the massive $U(1)$ is not 
directly visible in an F-theory compactification with only harmonic 
forms, since in the absence of fluxes all gaugings and D-terms disappear.
However, the gauging induced by \eqref{CSact} precisely maps to the 
gauging \eqref{gaugedF4} and hence identifies the correct massive $U(1)$
in F-theory. Clearly also the scalar $c$ in \eqref{C3_exp_c} is 
identified with the scalar arising in the R-R two-form $C_2$.
Note that this mechanism clarifies why in a Type IIB compactification 
a stack of $N$ branes not invariant under the orientifold action gives rise 
to gauge group $U(N)$,
 while in F-theory one generically sees only the $SU(N)$: The perturbative 
St\"uckelberg mechanism due to (\ref{CSact}) is built into the geometry 
automatically and thus does not allow us to disentangle the massive 
$U(1)$ boson from the other Kaluza-Klein states. 
However, linear combinations $F$ of such $U(1)$s may  remain massless as far as the \emph{geometric} St\"uckelberg is concerned, both in IIB and in F-theory.
In the presence of internal gauge flux $\langle F \rangle$ these remaining $U(1)$ potentials may still receive a mass term from the independent coupling 
\beq
 \label{mass2}
   S_{CS} \supset \int  F \wedge \langle F \rangle  \wedge C_4
\eeq
in IIB language. For a clean distinction between the two St\"uckelberg 
terms (\ref{CSact}) and (\ref{mass2}) in the recent Type IIB  literature see \cite{Plauschinn:2008yd}. 
The M/F-theory analogue is the Chern-Simons coupling
\beq
 \label{mass3}
   S =  \int  C_3  \wedge F_4 \wedge G_4,
\eeq 
where we distinguish the flux $G_4$ notationally from the field strength $F_4$ as defined in (\ref{gaugedF4}).
 Only the combinations $F$  of gauge fields  that lie also in the kernel of the mass matrix resulting from 
(\ref{mass2}) remain as true gauge symmetries.  Unlike the geometric mass terms due to couplings of the 
form (\ref{CSact}) the flux-induced mass term  turns out to lie below the Kaluza-Klein scale in F-theory. 
More details will be provided in \cite{progress1}.

\section{$U(1)$-restricted Tate models and dimension-4 proton decay} 
\label{U(1)-restricted}

\subsection{Split spectral covers and dimension-4 proton decay}
\label{SplitSCC}

The structure outlined in the previous section on the basis of the global Tate model is captured 
locally by the spectral cover construction. While eventually we aim at going beyond this local 
picture, we now recall its basic features. This makes contact with the F-theory GUT model building 
literature and allows us to analyse potential limitations of this technology.

The spectral cover approach to F-theory model building  \cite{Donagi:2008ca,Hayashi:2008ba,Donagi:2009ra,Hayashi:2009ge} 
can be applied in situations with non-abelian gauge symmetry along just a single divisor  $S: w=0$.
Its essence is to focus on the local neighbourhood of $S$ within $Y$
by discarding all terms of higher power in the normal coordinate $w$ that appear in the sections  $\mathfrak{b}_n$.
The restrictions of  $\mathfrak{b}_n$ to the GUT divisor,
\bea
\label{bi}
b_n = \mathfrak{b}_n|_{\omega=0},
\eea
are therefore sections entirely on $S$.
In this local picture,  the GUT brane is described as the base of the bundle $K_S \rightarrow S$, given by $s=0$.
 The neighbourhood of $S$ is then modelled by a spectral surface viewed as a divisor of the total space of $K_S$. In the sequel we will concentrate on the Tate model for an $SU(5)$ GUT symmetry along $S$ with associated spectral surface
 \beq \label{C5b}
\mathcal{C}^{(5)}:  \,  b_0 s^5 + b_2 s^3 + b_3 s^2 + b_4 s + b_5 = 0.
\eeq
One can think of $\mathcal{C}^{(5)}$  as encoding the information about the  discriminant locus in 
the local vicinity of $S$.
In particular, the intersections of $\mathcal{C}^{(5)}$ and $S$ determine the matter curves (\ref{curve10}) on $S$. 
It is also clear from the relation (\ref{bi}), though, that all the information in $\mathfrak b_n$ 
contained in  terms higher in $w$ is lost in the spectral cover approach.

In agreement with our previous remarks, the gauge group $G$ along $S$ arises by breaking an 
underlying $E_8$ gauge symmetry via a Higgs bundle of structure group $H={E_8}/G$. 
For $G=SU(5)_{GUT}$ this means that $H= SU(5)_{\perp}$.
Note that this interpretation of the gauge group $G$ was motivated in \cite{Donagi:2008ca,Hayashi:2008ba} 
by the unfolding of an $E_8$ symmetry that underlies the local gauge theory 
on $S$ (see also \cite{Beasley:2008kw,Beasley:2008dc}) and the spectral cover construction 
is a geometrisation of this local gauge theory.
As anticipated in section \ref{E8-Cartan} we will argue in section \ref{SCCs} that the 
interpretation of the gauge group $G$ along $S$ as the compliment of the spectral cover 
group  $H$ in $E_8$ arises naturally from the geometric Tate model. Our interpretation 
of the spectral cover makes no reference to a local gauge theory description let alone 
to a heterotic dual.

If $\mathcal{C}^{(5)}$ splits into two or more divisors, the structure group $H$ 
factorises accordingly. As a result, the gauge group $G = E_8/H$ is expected to increase. 
E.g. let us specify to the situation of a factorised divisor \cite{Marsano:2009gv}
\beq
  {\cal C}^{(5)} = {\cal C}^{(4)} \times {\cal C}^{(1)}.
\eeq
This split corresponds to the factorisation of \eqref{C5b} into
\beq \label{C4-1}
  (c_0 s^4 + c_1 s^3 + c_2 s^2 + c_3 s + c_4) (d_0 s + d_1) = 0.
\eeq
By comparison of \eqref{C4-1} with \eqref{C5b} one can express the sections $b_n$ as 
\beq \label{bcrel}
  b_5 = c_4, \quad b_4 = c_3 + c_4 d_0, \quad b_3 = c_2 + c_3 d_0, \qquad
  b_2 = c_1 + c_2 d_0, \quad b_0 =  -c_1 d_0^2\ ,
\eeq
where we have restricted to $ c_0 = -c_1 d_0$ such that the term proportional 
to $s^4$ vanishes in \eqref{C4-1}. 
All $b_n, c_n$ are appropriate sections on $S$ such that the Tate sections 
$a_n$  (\ref{TateSUn})  sections are elements of  $H^0(B,K_B^{-n})$, and $d_1$ 
is taken as a constant in order to avoid unwanted ${\bf 10}$ matter curves, 
see \cite{Marsano:2009gv} for details.

Since under this split $H=SU(4) \times U(1)_X$, the gauge group $G$ is expected to 
enhance to $SU(5) \times U(1)_X$ and the massless $SU(5)_{GUT}$ matter 
picks up the following $U(1)_X$ charges,
\beq
\label{charges}
{\bf 10}_1,\quad \quad {\bf 10}_{-4}, \quad\quad  
({\bf \ov 5}_m)_{-3}, \quad\quad ({\bf 5}_H)_{-2} +  ({\bf \ov 5}_H)_{2}, 
\eeq
where the exotic ${\bf 10}_{-4}$ is absent for the choice \eqref{bcrel}.
The different $U(1)_X$ charges of the ${\bf 5}$ representations reflect the split of 
the {\bf 5}-matter curves on $S$ which is induced by the factorisation of the spectral cover.

The charge assignments in the $S[U(4) \times U(1)_X]$  follow group theoretically 
from the decomposition of $SU(5)_{\perp}\rightarrow S[U(4)_{\perp} \times U(1)_X]$ 
by identifying the generator of $U(1)_X$ as the $SU(5)$ Cartan generator $T={\rm diag}(1,1,1,1,-4)$. 
Correspondingly, the representations of $SU(5)_{\perp}$ in the decomposition of the ${\bf 248}$ of $E_8$ into $SU(5) \times SU(5)_{\perp}$,
\beq \label{matter-decomp-SU(5)}
  {\bf 248} \mapsto ({\bf 24},1) + (1,{\bf 24}) + [ ({\bf 10}, {\bf 5}) + ({\bf \ov 5}, {\bf 10}) + h.c.],
\eeq
become
\beq
 {\bf  5} \rightarrow {\bf 4}_1 + 1_{-4}, \quad\quad {\bf 10} \rightarrow {\bf 6}_2 + {\bf 4}_{-3}  \quad\quad
  {\bf 24}_0 \rightarrow {\bf 15}_0 + 1 + {\bf 4}_{5} + {\bf \ov 4}_{-5} \label{group-split2}.
\eeq
At the level of roots and weights this can be phrased as follows:
The nodes  $\omega_i$ of the extended Dynkin diagram of $E_8$ are partitioned into 
the simple roots $\omega_i^G$ and $\omega_i^H$ with $G=SU(5)_{GUT}$ and $H=SU(5)_\perp$. 
The fundamental weights $\lambda_i, \, i=1, \ldots 5$ of $SU(5)_\perp$ have the well-known representation in terms of $\omega_i$
given in \eqref{fund_weights}. 
The elements of the Cartan subalgebra of $H$ are identified with the dual elements $\lambda_i^*$, introduced in \eqref{dual-weights}. 
In particular, the generator $T_X$ corresponds to the combination
\beq
\label{TX}
     \lambda_1^*  + \lambda_2^*  + \lambda_3^* +  \lambda_4^* - 4 \lambda_5^* .
\eeq
This guarantees the correct $U(1)_X$ charges for the above states. E.g.~the  
representation $(1,{\bf 4}_{5})$ under $SU(5)_{GUT} \times S[U(4) \times U(1)_X]$ that descends 
from   $(1,{\bf 24})$ corresponds to the weight $\lambda_i - \lambda_5$, $i=1, \ldots, 5$, and 
thus has $U(1)_X$ charge $1-(-4) =5$.
Treating all $\lambda_i$ independent as above is the local version of the Coulomb branch for $\ov Y_{E_8}$ 
advocated in section \ref{E8-Cartan}. The spectral cover analogue of the deformation to $\ov Y_{SU(5)}$ 
corresponds to the introduction of monodromies acting on the roots 
within $H$  \cite{Hayashi:2009ge,Bouchard:2009bu,Marsano:2009gv,Hayashi:2010zp}. In the split 
spectral cover of the type \eqref{C4-1}, the monodromies act not on the full $SU(5)_\perp$ by 
permutation of all $\lambda_i, i=1, \ldots,5$, but only on an $SU(4)$ subgroup of $SU(5)_\perp$ by 
identifying $\lambda_i$, $i=1,\ldots,4$. This implies that the $U(1)_X$ associated with the  
generator $T_X$ might have a chance to survive the Higgsing/deformation.

If the $U(1)_X$ factor really survives globally it leads to phenomenologically appealing 
selection rules for the Yukawa couplings of the GUT theory. In particular split spectral 
covers are used in the compact models of \cite{Marsano:2009ym, Blumenhagen:2009yv,Marsano:2009wr,Grimm:2009yu} 
as a method to avoid dimension-4 proton decay operators because couplings of the type
${\bf 10} \, {\bf \ov 5 }_m \, {\bf \ov 5}_m$ are forbidden while the desirable Yukawa 
coupling ${\bf 10} \, {\bf \ov 5 }_m \, {\bf \ov 5}_H$ is allowed by $U(1)_X$.

Another virtue of the spectral cover construction is that it yields  a description also of 
gauge flux \cite{Donagi:2008ca,Hayashi:2008ba,Donagi:2009ra}.
For a split spectral cover of the type (\ref{C4-1}) one can express a certain class of gauge 
flux in terms of an $S[U(4) \times U(1)_X]$ bundle $W$ \cite{Blumenhagen:2009yv} on $S$. 
This is a spectral cover bundle for which the spectral sheaf factorises into  ${\cal N}^{(4)}$ 
and  ${\cal N}^{(1)}$ defined respectively on ${\cal C}^{(4)}$ and ${\cal C}^{(1)}$. 
Its description involves an element $\zeta \subset H^2(S,\mathbb Z)$ such that 
$\pi_4^*{\cal N}^{(4)} = \zeta = - \pi_1^*{\cal N}^{(1)}$,  where $\pi_i$ is the 
projection ${\cal C}^{(i)} \rightarrow S$.
Following the logic of heterotic spectral cover constructions with $S[U(N) \times U(1)]$ 
bundles \cite{Blumenhagen:2006ux,Blumenhagen:2006wj} it is natural to assume a D-term 
potential for $U(1)_X$ of the standard form
\beq
\label{Dterm1}
\sum_i q_i |\phi_i|^2 + \xi =0 , \qquad \qquad  \xi \propto \int_S J \wedge \zeta.
\eeq
Here $\phi_i$ denotes charged matter under $U(1)_X$ and we have also displayed the 
Fayet-Iliopoulos D-term for the gauge bundle.
Associated with this D-term is a St\"uckelberg-type mass term for the $U(1)_X$ boson 
induced by non-zero gauge flux. The fact that the $U(1)_X$ boson acquires a St\"uckelberg 
mass is well-known not to affect its relevance as a global symmetry constraining the Yukawas.

However, an important caveat concerning the existence of $U(1)_X$ as a gauge symmetry, 
albeit a massive one, is that the local split of the spectral cover is by construction 
insensitive to information away from the GUT brane $S$. In particular there can be matter 
states localised away from $S$ which are uncharged under the non-abelian part of the GUT 
group $G$, but charged under $U(1)_X$.
In fact in $SU(5)_{GUT}$ models based on split spectral covers of the type (\ref{C4-1}) the role of right-handed neutrinos is played by states 
\beq
\label{NRc}
N_R^c:   \, \, 1_5.
\eeq
These are the states commented on after (\ref{TX}). They have the correct $U(1)_X$ quantum numbers to participate in the Dirac Yukawa coupling
${\bf \ov 5}_m \, {\bf 5}_H \, N_R^c$.
Since these states arise away from the GUT brane $S$, their precise location is hard 
to determine in the spectral cover approach, 
see \cite{Blumenhagen:2009yv,Marsano:2009wr}  for proposals.

 Now, the problem is that the \emph{local} split of the spectral cover does not 
guarantee that matter of the type (\ref{NRc}) does not acquire a non-zero vacuum 
expectation value (VEV) such as to \emph{higgs} $U(1)_X$. 
In fact, from a field theory perspective any VEV of such recombination moduli in 
agreement with the D-term condition (\ref{Dterm1}) is compatible with the local 
split (\ref{C4-1}), which only takes into account the vicinity of $S$.
 Turning tables around, the unambiguous appearance of \emph{localised} massless 
matter away from $S$ can be taken as an indication that $U(1)_X$ is un-higgsed. 
 To really determine the presence of such massless matter one has to go beyond 
the spectral cover approximation and consider the full Tate model. 
 The  matter in question must then be localised on a curve $C$ on the $I_1$ 
part of the discriminant over which the singularity type of the fiber enhances 
from $I_1$ to $I_2$. This corresponds to an enhancement to $A_1 \simeq SU(2)$ 
and extra matter will appear from the decomposition of the adjoint ${\bf 3}$ of $SU(2)$ under the branching $SU(2) \rightarrow U(1)$.

To appreciate the consequences of a higgsing of $U(1)_X$ for proton decay let us assume that indeed a recombination modulus $\Phi$  
localised away from $S$ has acquired a VEV that would not be detected from the \emph{local} split of the spectral cover.
Such a field could have the quantum numbers of $N_R^c$ with $U(1)_X$ charge $+5$ or of its conjugate with charge $-5$.
In the first case $\Phi$ could participate in a dimension-5 coupling
$W \supset  \frac{1}{M}\,\,  {{\bf 10} \, {\bf \ov 5}_m  \,  {\bf \ov 5}_m \,  {\bf \Phi} }$,
where $M$ is a mass scale. 
Clearly a VEV for  $\Phi$ induces a dimension-4 proton decay operator
\beq
\frac{\langle {\bf \Phi} \rangle }{M}\, {\bf 10} \,  {\bf \ov 5}_m  \,  {\bf \ov 5}_m  .
\eeq
If, on the other hand, it is the field ${\bf \tilde \Phi}$ with charge $-5$ that acquires a VEV 
this way of generating  dangerous $U(1)_X$ violating  dangerous couplings does not occur.
Which of the two scenarios arises depends on an interplay of the sign of the Fayet-Iliopoulos 
term and of further F-terms constraining the VEVs of ${\bf \Phi}$ and ${\bf \tilde \Phi}$.
For a discussion of similar effects in the 
heterotic context see \cite{Tatar:2006dc,Weigand:2006yj,Anderson:2010tc}.

A related danger for models with higgsed $U(1)_X$ concerns the precise counting of massless matter.
As stressed in this context in \cite{Blumenhagen:2009yv} (see also \cite{Tatar:2009jk}) a VEV of the 
recombination moduli is identical to a deformation of the $S[U(4) \times U(1)_X]$ bundle $W$  into a 
proper $SU(5)$ bundle. Loosely speaking this can be thought of as forming a non-split extension from 
a direct sum of bundles, even though in this context $W$ is actually not a direct sum of two independently defined vector bundles.
In any case, while in this process the total chirality of the model is unaffected, the chirality of individual 
matter species might change.
Concretely if the recombination modulus ${\bf \Phi}$  with charge $+5$ couples to ${ \bf \ov 5}_m$ and 
${\bf 5}_H$ as $ W \supset {\bf \Phi} \,  {\bf 5}_H \,   {\bf \ov 5}_m$ a VEV for  ${\bf \Phi}$ produces 
also a mass term of the form $\langle {\bf \Phi} \rangle {\bf 5}_H \,  {\bf \ov 5}_m$.  
Therefore in general the computation of the individual chiralities of models with factorised 
spectral cover is guaranteed to be valid only if $U(1)_X$ is un-higgsed.

\subsection{$U(1)$-restricted Tate models and un-higgsed $U(1)_X$    }
\label{Globalfac}

In this section we will introduce the geometries which guarantee the existence of an un-higgsed $U(1)_X$.
In fact we will see that the existence of the desired $U(1)_X$ symmetry can be ensured provided 
one extends the factorisation of the spectral cover to a global restriction of the full Tate model. 
In the following we will refer to these geometries as \textit{$U(1)$-restricted Tate models}. 
Let us promote the split \eqref{C4-1} to a \emph{global} modification of the sections 
$\mathfrak b_n$ as
\bea
\label{global-split}
&&     \fb_5 = \fc_4, \quad \fb_4 = \fc_3 + \fc_4 d_0, \quad \fb_3 = \fc_2 + \fc_3 d_0, \\  
&&  \fb_2 = \fc_1 + \fc_2 d_0, \quad  \fb_0 =  -\fc_1 d_0^2\ ,
\eea
where $\fb_n,\fc_n$ now depend on all coordinates of the base $B$.
With this form of the sections inserted into the Tate form \eqref{Tate1}, the coordinate transformation 
\bea
  x \rightarrow \tilde x + w^2 d_0^2  ,\quad y \rightarrow \tilde y - w^3 d_0^3\
\eea
brings the Tate model polynomial into the form
\beq \label{Tate2}
    P_{\rm T} = \tilde x^3 - \tilde y^2 +  \tilde x\,  \tilde y\,  \tilde a_1 +  \tilde x^2\, \tilde a_2 +  \tilde y\, \tilde a_3   +  \tilde x\, \tilde a_4 \ = 0 
\eeq
with
\bea \label{TateSU5b1}
  \tilde a_1 &=& \tilde \fb_5 \ \ = \fc_4 ,  \nn \\
  \tilde a_2 &=& \tilde \fb_4 w\ = (\fc_3 + \fc_4 d_0 + 3 w d_0^2) w , \\
  \tilde a_3 &=& \tilde \fb_3 w^2= (\fc_2 + \fc_3 d_0 + \fc_4 d_0^2 + 2 d_0^3 w) w^2 , \nn  \\
  \tilde a_4 &=& \tilde \fb_2 w^3 = (\fc_1+\fc_2 d_0 +2 \fc_3 d_0^2+ \fc_4 d_0^3 + 3 w d_0^4) w^3 . \nn 
 \eea
Crucially one notes that $\tilde a_6 = \tilde \fb_0 = 0$ and that the coefficients $\tilde \fb_n$ are generic since the $\fc_n$ are 
generic. We denote the resulting singular fourfold by $X_{SU(5)}$, or, for more general gauge groups over $S$, by $X_{G}$. 
Note that on $X_{SU(5)}$ the ${\bf 5}$ curve given in \eqref{curve10} now splits as 
\beq
    \tilde P_{\bf 5} = \tilde \fb_3 ( \tilde \fb_4 - \tilde \fb_2 \tilde \fb_5) = 0, 
\eeq 
which allows for the localisation ${\bf 5}_m$ and ${\bf 5}_{H}+{\bf \ov 5}_{H}$ on different curves.
Moreover, the discriminant of $X_{SU(5)}$ is now of the form 
\bea
  \Delta &=& w^5 \Big( \tilde \fb_2 \tilde \fb_3 \tilde \fb_5 \big[\tilde \fb_5^4 + 8\tilde \fb_4 \tilde \fb_5^2 w + 16 w^2 (\tilde \fb_4^2 - 6 \tilde \fb_2 w)\big]
                       + \tilde \fb_3^3 \tilde \fb_5 w (\tilde \fb_5^2 + 36\tilde  \fb_4 w)  \nn \\ 
&& \phantom{w^5 \big(} - \tilde \fb_3^2 \big[\tilde \fb_4 \tilde \fb_5^4 + 8 \tilde \fb_4^2 \tilde \fb_5^2 w + 2 (8 \tilde \fb_4^3 + 15 \tilde \fb_2 \tilde \fb_5^2) w^2  
 - 72 \tilde \fb_2 \tilde \fb_4 w^3 \big]  \\
  &&\phantom{w^5 \big(} +  \tilde \fb_2^2 w \big[\tilde \fb_5^4 + 8\tilde \fb_4 \tilde \fb_5^2 w + 16 w^2 (\tilde \fb_4^2 - 4\tilde \fb_2 w) \big] - 27 \tilde \fb_3^4 w^3  \Big)\ . \nn
\eea
This implies that the elliptic fibration is singular over the curve 
\beq \label{sing_curve}
  C : \qquad \tilde \fb_2 = 0 \ ,\qquad \tilde \fb_3 = 0 
\eeq
in the base $B$. Application of Tate's algorithm confirms an $SU(2)$ enhancement with $\tilde a_n$ weights 
$(0,0,1,1,2)$ and $\Delta$-weight $2$ along $C$. 
This singular curve describes the self-intersection locus of the $I_1$ part of the discriminant appearing in the brackets.
Here extra massless degrees of freedom appear in the singular limit according 
to the decomposition of the adjoint of $SU(2)$. The $U(1)_X$ charge of the massless states along $C$ derive from the specific 
embedding into $E_8$ and will be scrutinised further in the next section.
We thus interpret the specialisation to (\ref{TateSU5b1}) as the un-higgsing 
of the $U(1)_X$ gauge symmetry, signalled by the appearance of massless charged matter. 
These charged massless degrees of freedom play the role of the recombination 
moduli whose VEV in turn  smoothens out the singularity away from the locus (\ref{TateSU5b1}).
This intuitive picture will be corroborated further below for a simple F-theory 
model with orientifold limit. Note that despite the appearance of a singular 
self-intersection curve $C$ the full $I_1$ piece does not factorise. 
However, this is not required for the existence of the abelian $U(1)_X$ 
gauge boson. 

Note that this analysis generalises to other gauge groups $G$ localised on the divisor $S$, 
as long as $G\subset E_7$. 
Due to the additional $U(1)$ factor the maximal gauge group $E_8$ is no longer attainable 
along a divisor. This does not mean, though, that higher codimension $E_8$ enhancements 
along curves or points are forbidden. In fact, points of $E_8$ enhancement have been 
argued to lead to a favourable flavour structure in \cite{Heckman:2009mn}.

\subsection{Abelian gauge bosons from resolution and connection to $E_7$ fibers}
\label{Bosons}

So far we have motivated the existence of an un-higgsed 
$U(1)_X$ by slightly indirect  methods. 
A direct argument is via the logic described 
in section \ref{E8-Cartan}, where we described how 
in M-theory on a Calabi-Yau fourfold one can study $U(1)$ 
gauge factors by moving to the Coulomb branch of the gauge theory. 

Applying the general relation \eqref{nv} to the Tate model \eqref{Tate2} one first encounters
four $U(1)$ factors which correspond to the 
Cartan generators of $G=SU(5)_{GUT}$ and which arise from
blow-up divisors $D^G_i$ with Poincar\'e dual two-forms $\omega^G_i$.
These have been explicitly constructed for GUT models 
in refs.~\cite{Blumenhagen:2009yv,Grimm:2009yu}.
Crucially, here we will encounter an additional $U(1)_X$, since $Y$ also contains
the singular curve $C$, given in \eqref{sing_curve}, 
away from the GUT-brane $S$. The singular curve can be 
canonically resolved into a divisor $\hat D_C$ by introducing 
a new coordinate $s$ and replacing 
$\tilde y \rightarrow \tilde ys,\, \tilde x \rightarrow \tilde xs$. 
$\hat D_C$ is then given by $s=0$ and increases the number $h^{1,1}(\ov Y)$
in \eqref{nv} by one. We denote the Poincar\'e dual two-form by $\omega_C$. 
One checks that $\ov Y$ is indeed non-singular 
after appropriately introducing scaling relations for $s$. 
Due to the extra element in $H^{1,1}(\ov Y, \mathbb Z)$ there must exist a harmonic two-form $\omega_X$ which is related to $\omega_C$
and which encodes the surviving $U(1)_X$ factor. Simple examples show that a candidate for 
$\omega_X$ is 
\beq \label{def-omegaX}
   \omega_X = \omega_C - \omega_B + \pi^{*} c_1(B)\ ,
\eeq
where $\pi$ is the map from $X_G$ to its base $B$.
Using this two-form the expansion for $C_3$ reads 
\beq \label{C3exp_2}
C_3 = A_X \wedge \omega_X + \sum_i A^i \wedge \omega^G_i + \ldots\ ,
\eeq
where $A_X$ is the $U(1)_X$ potential and the $A^i$ correspond to 
the Cartan $U(1)$s in the non-abelian gauge group. Note that one again 
has to define linear combinations of the $A^i$ and the  $\omega^G_i$, associated with the 
simple roots, to obtain the Cartan $U(1)$s. 
Moving to the singular $Y_G$, the $A^i$ will become part of the 
non-abelian gauge group over $S$. However, since $Y$ away from the GUT-brane 
did not become singular over a divisor, but over a curve $C$ there is 
no non-abelian gauge enhancement involving the $A_X$ factor, and the $U(1)_X$ remains untouched. 
This establishes the appearance of the $U(1)_X$ gauge boson.

Let us now resolve the $SU(2)$ singularity along $C$ explicitly. For 
simplicity we will only concentrate on 
$C$ and refer for resolution of the $SU(5)$ singularity along the 
GUT brane $S$ to refs.~\cite{Blumenhagen:2009yv,Grimm:2009yu}.
The resolved Tate form is given by
\beq \label{Tate2_res}
    P_{\rm T} = \tilde x^3 s^2 - \tilde y^2 s + \tilde x\, \tilde y\,  \tilde z\, a_1 
       + s\, \tilde x^2\, \tilde z^2\, a_2 + \tilde y\, \tilde z^3\,a_3   + \tilde x\, \tilde z^4\, a_4 = 0\ .
\eeq 
The new Calabi-Yau $\ov X_G$ manifold 
has two fibers: Setting $\tilde a_n=1=s$ we recover the original ${\mathbb P}_{1,2,3}[6]$ fiber, while for  $\tilde a_n=1=\tilde x$ one finds a so-called $E_7$ fiber.
The latter is given by the hypersurface  ${\mathbb P}_{1,1,2}[4]$ in the ambient space  $(\tilde y,\tilde z,s) \cong (\lambda y,\lambda z,\lambda^2 s)$.
This fiber has 
two sections. Note that the section $z=0$ is shared by the new 
${\mathbb P}_{1,1,2}[4]$ fiber as well as the original ${\mathbb P}_{1,2,3}[6]$ fiber, and
yields the base $B$. In six-dimensional F-theory compactifications 
such geometries have been studied in the context of heterotic/F-theory duality in refs.~\cite{Candelas:1997pv,Berglund:1998va}. 
Note that the resolution \eqref{Tate2_res} can be performed for all gauge groups $G \subset E_7$ on 
$S$. It turns out that enhancement to a gauge group $E_8$ is no longer possible on the divisor $S$. Clearly, this fits with the 
group theory interpretation presented at the end of section \ref{Globalfac}, since $E_7 \times U(1)$ is of maximal rank in $E_8$.
In other words the diagram \eqref{rec-tab} gets now replaced by 
\beq
\label{rec-tab_2}
    \begin{array}{lcl}X_{E_7 \times U(1)} & \xrightarrow{\ \ \text{brane recomb.}\ \ } &   X_{G \times U(1)} \\
                      \downarrow &  & \downarrow \\
                  \ov X_{E_7 \times U(1)} & \xrightarrow{\ \ \text{geom.~transition}\ \ }  & \ov X_{G \times U(1)} \end{array}
\eeq
where we have included the $U(1)$ factor to stress the difference with the geometries in \eqref{rec-tab}.
In section \ref{GlobalSCCs} we will show that the split spectral cover group and its splittings 
can be studied by using mirror symmetry for $\ov X_{G}$ and a reference geometry $\ov X_{E_7}$.

Since the underlying gauge group for the $U(1)$-restricted Tate model is not $E_8$, but 
rather $E_7 \times U(1)$, it is natural to ask how the abelian gauge group in the final model 
embeds into $E_8$. To demonstrate this 
let us specialise to $G=SU(5)$. In this case we expect the surviving abelian group to be 
given by $U(1)_X$ from the breaking $E_8 \rightarrow SU(5) \times SU(4) \times U(1)_X$. 
That this is indeed the case can be seen as follows:
The abelian gauge group visible in the underlying $X_{E_7}$ is the up to normalisation unique Cartan $U(1)$ within $E_8$ responsible for the branching
\beq
E_8 \rightarrow E_7 \times U(1)_a, \quad\quad\quad
{\bf 248} \rightarrow {\bf 133}_0 + {\bf 56}_1 +  {\bf \ov{56}}_{-1} + 1_2 + 1_{-2}. \nonumber
\eeq
The deformation to $X_{SU(5)}$ can be understood via the branching $E_7 \rightarrow SU(5)\times SU(3)_\perp \times U(1)_b$.
The full branching rules to $SU(5) \times SU(3)_\perp \times U(1)_a \times U(1)_b$ are
\bea
&&  {\bf 133}_0 \rightarrow (1,{\bf 8})_{0,0} + ({\bf 24},1)_{0,0} + (1,1)_{0,0} + [({\bf 5},1)_{0,6} + ({\bf 5},{\bf 3})_{0,-4} + ({\bf 10},{\bf 3})_{0,2} + c.c ], \nn \\
&&  {\bf 56}_1 \rightarrow (1,{\bf 3})_{1,-5} + ({\bf 5},{\bf 3})_{1,1} + ({\bf 10},1)_{1,-3} + c.c., \quad\quad\quad  1_2 \rightarrow 1_{2,0},
 \eea
with the subscripts denoting the charges $(q_a, q_b)$. 
The deformation to the $U(1)$-restricted, but otherwise most generic Tate model (\ref{global-split}) 
higgses $SU(3)_\perp$ and one linear combination of $U(1)_a$ and $U(1)_b$.
The remaining $U(1)$ symmetry can be determined by noting that the model contains only one matter 
curve on which an $SU(5)$ singlet localises. Recall that this is precisely the curve $C$ appearing 
in (\ref{sing_curve}). Indeed the up to rescaling unique combination of $U(1)_a$ and $U(1)_b$ 
compatible with this is 
\beq \label{U(1)comb}
U(1)_X = \frac12 \Bigl( -5 \,  U(1)_a  +  U(1)_b   \Bigr),
\eeq
which results in the anticipated $U(1)_X$ charges displayed in  (\ref{charges}). Note that the 
surviving $U(1)_X$ generator embeds indirectly into $E_8$ via the branchings and higgsing described above.

The $U(1)$ restriction of the Tate model has another important effect: The Euler characteristic 
of the resolved fourfold $\ov X_G$ decreases considerably compared to the original fibration.
This can be traced back to the fact that the $U(1)$ restriction \eqref{Tate2}, \eqref{Tate2_res} forces us to 
fix many complex structure moduli to restrict the $I_1$-locus.  
The  change in the Euler characteristic can be given as a closed expression if one restricts 
to the case of a smooth Tate model $Y$ with no non-abelian enhancement. 
The Euler characteristic of the fourfold can be computed as \cite{Sethi:1996es,Klemm:1996ts}
\beq
\label{4-fold_chi}
\chi(Y) = 12 \int_{B} c_1(B) \, c_2(B) + 360 \int_{B} c_1^3(B).
\eeq
After the $U(1)$ restriction and resolution the Euler characteristic is reduced as
\beq
 \chi(\ov X) = \chi(Y) - 216 \int_{B} c_1^3(B),
\eeq
which corresponds to the value obtained for an $E_7$ fibration.  This can be checked 
explicitly in examples with the help of toric geometry. Using the general 
formula $\chi = 6 (8 + h^{1,1} + h^{3,1} - h^{2,1})$ one determines straightforwardly 
the number of complex structure moduli which have to be fixed in  order to ensure the 
presence of the addition $U(1)$.
For the phenomenologically interesting cases of $SU(5)$ GUT models the
 effect of the $U(1)$ restriction has to be computed by direct resolution. 
For instance, consider the 3-generation model in the main text of \cite{Grimm:2009yu}: 
The Tate model corresponding to the \emph{non-split} $SU(5)_{\perp}$  spectral cover 
(\ref{C5b}) gives rise to  $\chi=5718$. Once we implement the global $U(1)$ restriction 
(\ref{global-split}) we find instead $\chi=2556$ after resolution. On the other hand,
 if one computes the value of $\chi$ based just on the split spectral cover (\ref{bcrel}) 
as opposed to the globally $U(1)$-restricted Tate model (\ref{global-split}), one finds 
$\chi=5424$ using the formula of \cite{ Blumenhagen:2009yv}. This shows that promoting 
the split spectral cover to a global $U(1)$-restricted Tate model decreases the value of $\chi$ significantly. 
Note that this affects all previous models ~\cite{Marsano:2009gv, Blumenhagen:2009yv,Marsano:2009ym,Grimm:2009yu}  
based on split spectral covers in the literature and makes a re-evaluation of the D3-tadpole condition necessary 
if one wants to promote the split globally to save the $U(1)$ selection rules.  This demonstrates once more the 
entanglement of global properties of the model and the appearance of abelian symmetry and selection rules.

We conclude this section with some remarks on  the description of $U(1)_X$ gauge flux in 
$U(1)$-restricted Tate models of  type (\ref{Tate2}).
Recall from section \ref{E8-Cartan} that for a generic deformation from $\ov Y_{E_8}$ to  $\ov Y_{G}$ 
the Cartan flux associated with $H=E_8/G$ turns into data describing a non-abelian $H$-bundle. Such flux 
is represented by elements of $H^{2,2}(\ov Y_G)$ which cannot be written as the wedge of two two-forms. 
This is in agreement with the absence of a Fayet-Iliopoulos D-term because for generic deformations 
all $U(1)$ symmetries are higgsed.  For the $U(1)$ restricted Tate model, by contrast, due to the appearance 
of an unhiggsed $U(1)_X$ potential we do expect the presence of a Fayet-Iliopoulos term.  There must 
therefore exist a special type of $U(1)_X$ flux which  involves a truly abelian component. 
This is the global analogue of the extra $U(1)_X$ flux described by the class $\zeta \in H^2(S,\mathbb Z)$ 
within the split spectral cover approach, see the discussion around (\ref{Dterm1}).
In M-theory the D-terms arise from the Chern-Simons coupling
\bea
S_{CS} = - \frac{1}{12}\int C_3 \wedge G_4 \wedge G_4.
\eea
In terms of the element $\omega_X$ associated with the $U(1)_X$ generator the Fayet-Iliopoulos term is
\bea
\xi \propto \int \omega_X \wedge J \wedge G_4 .
\eea
This is the global version of the expected $U(1)_X$ Fayet-Iliopoulos term in the split spectral cover picture.
We leave it for future work to study the concrete description of this type of abelian gauge flux \cite{progress1}.

\subsection{Connection to brane recombination in orientifolds}
\label{Orientifolds}

We would like to end this discussion by giving yet another, complimentary argument for the 
appearance of an abelian gauge factor in $U(1)$ restricted Tate models of the form (\ref{Tate2}). 
This argument makes contact with the weakly coupled Type IIB orientifold picture as follows: 
For weakly coupled models, the restriction to $a_6=0$ describes the split of an orientifold invariant 7-brane into a brane-image brane pair.
To see this we need to recall that for a Weierstrass model (\ref{Weierstrass})
the connection with the IIB picture arises by the well-known 
Sen limit \cite{Sen:1997gv}. One parametrises without loss of generality 
\bea \label{def-fg}
   f = -3 h^2 + \epsilon \eta\ , \qquad
   g = -2 h^3+ \epsilon h\, \eta - \epsilon^2 \chi /12\ . 
\eea 
For the Tate model, which is related to the Weierstrass model via (\ref{fg}) and (\ref{beta}), one identifies
\bea
\label{weakTate}
\beta_2 = - 12 \, h, \quad  \beta_4= 2 \epsilon \,  \eta , \quad \beta_6= - \frac{1}{3} \epsilon^2\, \chi .
\eea
The orientifold limit consists in taking $\epsilon \rightarrow 0$ such that
the string coupling becomes weak away from $h=0$.
The leading order discriminant then reads
\beq
  \Delta_\epsilon = - 9 \epsilon^2 h^2 (\eta^2 - h \chi)  + {\cal O}(\epsilon^3)\, .
\eeq
For generic $\eta$ and $\chi$, corresponding to a smooth Weierstrass model without non-abelian gauge symmetries,
one  identifies one D7-brane and one O7-plane located at 
\beq
  \text{O7:} \quad h = 0 \ , \qquad \quad \text{D7:}\quad \eta^2 = h \chi.
\eeq
On the Calabi-Yau threefold double cover $Z: \xi^2 = h$ of the base space $B=Z/\mathbb Z_2$ one 
finds only one set of 7-branes \cite{Braun:2008ua,Collinucci:2008pf} wrapping the invariant 
cycle $Q: \eta^2 - \xi^2 \chi =0$. Correspondingly the abelian gauge boson is projected out by the orientifold action.

Let us now specialise the complex structure moduli such that the orientifold invariant brane along $Q$ 
splits into a  brane-image brane pair. As discussed in \cite{Collinucci:2008pf} this requires that
\bea
\label{deg1}
  \chi = \psi^2\ ,
\eea
which induces, again on the double cover Calabi-Yau $Z$, the split $Q \rightarrow Q_+ \cup Q_-$ 
with $Q_{\pm}: \eta \mp \xi \psi =0$. The orientifold action $\xi \rightarrow - \xi$ exchanges $Q_+$ and $Q_-$.

In view of the identification (\ref{weakTate}) together with the 
relation $\beta_6 = a_3^2 + 4 a_6$ given in (\ref{beta})  the 
factorisation (\ref{deg1}) means nothing other than $a_6=0$. 
Thus, the Tate model corresponding to the split brane-image 
brane pair is exactly of the $U(1)$ restricted form (\ref{Tate2}).
This orientifold picture makes the appearance of an extra $U(1)$ 
gauge symmetry clear: Prior to orientifolding the brane-image brane 
pair $Q_+ \cup Q_-$ carries gauge group $U(1)_+ \times U(1)_-$. 
The orientifold action identifies the  two factors such that a 
single  $U(1)$ boson survives corresponding to $U(1)_+ -  U(1)_-$. This is the $U(1)$ boson observed for 
restricted Tate models of the form (\ref{Tate2}). 
Note the crucial fact that for the simple model (\ref{deg1}) the branes $Q_+$ and $Q_-$ lie in the same homology class on $Z$. 
Therefore the geometric mass term discussed around (\ref{CSact}) does not make the 
linear combination   $U(1)_+ -  U(1)_-$ massive in agreement with the appearance of a $U(1)$ boson.
In the generic 
situation, however, the factorisation $Q \rightarrow Q_+ \cup Q_-$ 
is lost because of a deformation of $\chi = \psi^2$ into a 
non-factored form. This simply describes the higgsing of 
the $U(1)$ symmetry, whereby the extra matter states localised 
on the curve $C$ of $A_1$ singularities acquire a VEV.
From a more technical perspective, our analysis illustrates the 
connection between brane-image brane pairs and the appearance 
of restricted fibers in the Tate model, here fibers of $E_7$-type, 
see the discussion around (\ref{Tate2_res})\footnote{This connection was observed by the authors of 
the present article during completion of \cite{Blumenhagen:2009yv}. For an independent analysis see \cite{Aluffi:2009tm}.}.

Note that away from the strict orientifold limit $\epsilon \rightarrow 0$ the terms in the discriminant 
of higher order in $\epsilon$ become important. Taking them into account, the discriminant no longer 
factorises in O-plane and brane-image brane, but becomes a single component. This process is to be 
interpreted as the non-perturbative recombination of the brane and the O-plane system. 
However, it does not affect the presence of the abelian gauge boson. For recent advances in the context of weak-coupling 
Type IIB vs.~F-theory models see \cite{Blumenhagen:2008zz}, \cite{Collinucci:2008zs}-\cite{Cveti{c}:2010rq}.

\section{Global spectral covers and mirror symmetry}

\label{GlobalSCC}

\subsection{Spectral cover constructions}
\label{SCCs}

In this section we revisit the 
general philosophy behind the spectral cover approach to F-theory models. 
In particular we will argue for the appearance of a spectral cover directly 
from the form of the (globally defined) Tate model.
 
As reviewed, the general idea of the spectral cover is to describe the 
gauge group $G$ along a divisor $S$ by unfolding an underlying $E_8$ symmetry.
This picture arose in the description of local ALE fibrations 
over $S$  \cite{Donagi:2008ca,Hayashi:2008ba,Donagi:2009ra,Hayashi:2009ge}. Formally the same structure appears as in 
F-theory examples with a perturbative 
heterotic dual description  \cite{Friedman:1997yq}. 
In these cases the four-dimensional 
F-theory gauge group can be understood as the commutant of the structure group of a vector bundle embedded into the 
perturbative heterotic $E_8 \times E_8$. The heterotic vector bundle can be directly determined 
from the constraint of the F-theory manifold \cite{Morrison:1996pp,Berglund:1998ej}.
In the sequel we find more evidence for the relevance of  an $E_8$ bundle $V$ in the description of the four-dimensional 
gauge dynamics along $S$ in a genuine F-theory compactification of the type described in section \ref{fiber-restriction}.
Our considerations rely entirely on the structure of the Tate model without any reference to the local gauge dynamics on $S$ or a heterotic dual.
We therefore believe that this view sheds new light on the relevance of the spectral cover that helps understand also its role in compact models.

We consider F-theory compactifications
on elliptically fibered Calabi-Yau fourfolds
as introduced in section \ref{Tateform} and with a 
gauge enhancement $G$ over a \emph{single} divisor $S$ 
given by the constraint $w=0$. In a Tate model with $E_8$ elliptic 
fiber the generic fiber is given by $\mathbb{P}_{1,2,3}[6]$. 
As we will see in this case it is natural to consider gauge 
groups $G$ \emph{contained in} $E_8$. 
Given a $\mathbb{P}_{1,2,3}[6]$ fibration there is a natural split 
of the Tate constraint \eqref{Tate1} as 
\beq \label{PW-split}
   P_{\rm T} = P_{0} + P_{V}=0.
\eeq
It turns out that $P_V$ specifies a gauge bundle $V$ with structure group 
$H$ which breaks $E_8$ to its commutant $G=E_8/H$.
The simplest case is an $E_8$ 
singularity over $S$ corresponding to the Tate form
\bea \label{E8-Tate}
    P_0 &=& x^3 - y^2 + x\, y\,  z\, w h_6 + x^2\, z^2\, w^2 h_4 + y\, z^3\, w^3 h_3
     + x\, z^4\, w^4 h_2 +  z^6\, w^6 h_0, \nonumber\\
 P_{V} &=& z^6\, w^5 b_0.
\eea
Note that  $b_0$ can be chosen to be independent of $w$ 
by absorbing all higher order dependence on $w$ into $h_0$. 
Such a singular $Y$ can be constructed by studying 
the \textit{resolved} fourfold $\ov Y_{E_8}$ in which a set of resolution $\bbP^1$'s is 
fibered over $S$. 

Exactly in the case of an $E_8$ gauge group the leading 
powers of $w$ and $z$ match in $P_0$. One thus can introduce local coordinates $\tilde w = w z$ and $v = z^5 w^4 $
and write \eqref{E8-Tate} as
\bea
   P_0 &=& x^3 - y^2 + x\, y\, \tilde w h_6 + x^2\tilde w^2\, h_4 + y\,  h_3 
       + x\tilde w^3 h_2 +\tilde w^6 h_0, \nonumber  \\
   P_{V} &=& v\,\tilde w\, b_0 .
\eea
The point is that  $P_{V}$ is the defining equation for an $SU(1)$ bundle in the 
sense of the spectral cover as introduced in \cite{Friedman:1997yq}.
Note that the coordinate redefinition will generally induce inverse powers 
of $v$ in $P_0$, and hence is only valid in the patch of non-vanishing $v$. 
Before turning to a more global analysis, let us first focus on $P_V$. 
To reduce the gauge group from $E_8$ to a subgroup $G$   
one can systematically add new terms to $P_{V}$ which lower the 
vanishing orders $\kappa_n$ in \eqref{TateSUn}. This implies the following interpretation of the Tate model: The groups $G$ on $S$ are 
obtained as the \textit{deformation} of the original $E_8$ singularity 
in \eqref{E8-Tate} by allowing for new monomials 
with lower powers in $w$.  This introduces new complex structure deformations 
of the Calabi-Yau fourfold so as to change $P_V$ while keeping $P_0$ unaltered. In absence of a topological obstruction
enforcing a minimal gauge group, this process can be performed until all non-abelian 
gauge symmetry has been higgsed.
For example,  using the Tate formalism as in \cite{Bershadsky:1996nh} one finds the $P_V$ listed in table \ref{spec_table}.
\begin{table}[h!]
{\small
\beq \label{bundle_expressions}
  \begin{array}{c|c|l} G & E_8/G  & P_V \\
  \hline
      E_8 & SU(1) & v \tilde w b_0 \\
     E_7 & SU(2) & v(\tilde w^2 b_0 + x b_2) \\
     E_6 & SU(3) & v (\tilde w^2 b_0 + x \tilde w b_2 + y b_3)\\
     SO(10) & SU(4) &  v  (\tilde w^4 b_0 + x \tilde w^2  b_2 + y \tilde w  b_3 + x^2 b_4 ) \\
     SU(5) & SU(5) &  v  (\tilde w^5 b_0 + x \tilde w^3  b_2 + y \tilde w^2  b_3 + x^2 \tilde w b_4 + xy b_6 ) \\
     SU(4) & SO(10) & v ( \tilde w^2 y b_3 + \tilde w ^5 b_{0,2} + y x b_6 + \tilde w x^2 b_4      ) \\
                    && + v^2 ( \tilde w ^4 b_{0,1} + \tilde w ^2 x b_2  )
  \end{array} \nonumber
\eeq}
\caption{Spectral covers $P_V$ and their generalisations.} 
\label{spec_table}
\end{table}

In table \ref{spec_table} we have performed a coordinate redefinition to $\tilde w = z w$ and $ v= z^{6-N} w^{5-N}, \, N=1,\ldots,5$
to bring $P_V$ into the form of a spectral cover for $SU(N)$ bundles \cite{Friedman:1997yq}.
For $SO(10)$ bundles one redefines $v = z, \tilde w = z w$ 
and thus captures the terms  $z^6 ( b_{0,1} w^4 + b_{0,2} w^5)$ 
in the Tate form \eqref{E8-Tate} for an $SU(4)$ singularity.
As the polynomial 
for the $SO(10)$ bundle has a term $v^2$, such a bundle cannot be constructed directly  
by a spectral cover \cite{Friedman:1997yq}. However, the data of this 
bundle are encoded by the generalised spectral cover with higher powers of $v$. 
A more complete list including various other bundle 
groups can be found in ref.~\cite{Berglund:1998ej}.

Let us analyse the $SU(N)$ spectral covers in more detail by 
specifying the transformation of the coordinates  
$(v,\tilde w,x,y)$ and the $b_n$ as sections of appropriate  line bundles. 
Recall that $(z,x,y)$ appearing in (\ref{Tate1}) are sections 
$z \in H^0({\cal L})$, $x \in H^0({\cal L}^2 \otimes K_B^{-2})$, 
$y \in H^0({\cal L}^3 \otimes K_B^{-3})$, where $\cL$ is the line 
bundle for the scaling of $\mathbb P_{1,2,3}$, the ambient space of the 
elliptic fiber.
By definition $w$ is a section of $N_{S/B}$, 
the normal bundle to the divisor $S$ of $B$ over which we engineer non-abelian gauge enhancement.
With the above definition $\tilde w = z w$ and $ v= z^{6-N} w^{5-N}$ 
one arrives at $\tilde w \in H^0({\cal L}\otimes N_{S/B}  )$, $v \in H^0({\cal L}^{6-N}\otimes N_{S/B}^{5-N} )$.
Homogeneity of the polynomial $P_{\rm T}$ therefore uniquely determines the coefficients $b_n$ as sections
\beq
b_n \in H^0(S, \eta - n c_1(S)), \qquad \eta = 6 c_1(S) + c_1(N_{S/B}).
\eeq
This uses the adjunction formula
$K_B|_S = K_S \otimes N_{S/B}^{-1}$ as well as the fact that $b_n$ are truly sections of $S$ 
since all further dependence on $w$ has been shifted to $h_n$.
Recall that the construction of an $SU(N)$ bundle over $S$  
via spectral covers involves a spectral surface of class 
$N \sigma + \eta$, with $\sigma$ the section corresponding to $S$.
In the presented construction one recovers the spectral cover with 
$\eta$ from the geometry of $Y$ via $P_V$. 

It is crucial to keep in mind that the split \eqref{PW-split} was only 
possible because we assumed that the non-abelian gauge symmetry $G$  appears over 
the single divisor $S$ and that $G \subset E_8$. Despite this restriction, the geometry can be general
and no reference to the existence of a heterotic dual
has to be made. The non-trivial global information about the Calabi-Yau 
fourfold is captured by the sections $h_n$ in $P_0$, which are given 
already in the Calabi-Yau fourfold $Y_{E_8}$ with $E_8$ singularity 
\eqref{E8-Tate} with trivial $P_V$. One consequence of this 
construction appears to be the existence of 
a simple formula for the Euler characteristic of the resolved 
fourfold $\overline{Y}_G$ as \cite{Blumenhagen:2009yv}
\beq \label{chi-formula}
  \chi(\overline{Y}_{G}) = \chi(\overline{Y}_{E_8}) + \chi_{V}\ .
\eeq
Here $\chi_{V}$ is determined in a trivial 
fashion from the second Chern class of $V$ and can be 
computed for various bundles $V$ as a function of $\eta$ and the 
Chern classes of $S$ using \cite{Friedman:1997yq}. For example, 
for a vector bundle $V$ with structure group $SU(N)$  
one has 
\beq
   \chi_{V}^{SU(N)} = \int_S  c_1(S)^2 (N^3-N)+3N \eta (\eta-N c_1(S))\ .
\eeq
Note that there is no reason for \eqref{chi-formula} to be generally valid. 
Rather one should compare the Euler characteristic computed directly for 
an explicitly constructed and resolved Calabi-Yau fourfold $\ov Y_G$ 
with the value \eqref{chi-formula}.
A match indicates the global applicability of the spectral cover formalism.  
Such matches have been found explicitly for many examples~\cite{Blumenhagen:2009yv,Grimm:2009yu}.

\subsection{Mirror symmetry and spectral covers \label{mirror_spec}}

In the study of F-theory compactifications with 
non-abelian gauge symmetry one can  use 
two techniques to analyse the gauge sector. The first method is to study singularity enhancements 
over the divisor $S$. At each co-dimension the 
singularity can enhance further as
\beq \label{group-enhance}
   G \ \subset \ G_{C}\ \subset\ G_{P},
\eeq
where $C$ is an intersection curve of $S$ with the 
$I_1$ locus and $P$ is a point of intersection 
of $C$ with other $I_1$ curves in $S$. At the enhancement 
loci new matter fields and couplings can localise. Much of the 
information about the singularity is encoded in the 
canonical resolution $\ov Y_G$ by gluing in resolving $\bbP^1$s into 
the singular fibers over $S$, curves and points. In particular, 
the group enhancements are captured by the fact that 
the $\bbP^1$s intersect as the Dynkin diagrams of $G,G_C,G_P$
at the various locations in $S$ (see, e.g.~\cite{Intriligator:1997pq}).  
The second method to describe non-abelian enhancements 
is the generalisation of the constructions of \cite{Friedman:1997yq}
as described in section \ref{SCCs}. Here the situation is 
somewhat inverse to \eqref{group-enhance} since the bundles
breaking the $E_8$ become more trivial over curves and 
points. Thus the structure group reduces as 
\beq \label{H-groups}
  H\ \supset   \ H_C\ \supset\ H_P\ ,
\eeq
where $H= E_8/G,H_C=E_8/G_{C}$ and $H_P= E_8/G_{P}$ are the
respective commutants.
We have already stressed that this construction is much less 
general. In particular, 
only the analysis of F-theory compactifications with single 
groups $G$ in $E_8$ has been carried out. It would be 
interesting to explore generalizations of this construction.

In order to get deeper insights into the global applicability 
of the spectral cover construction and its extensions described 
in section \ref{SCCs} one can attempt to make $H,H_C,H_P$ visible 
as a physical gauge group in a dual theory. In ref.~\cite{Berglund:1998ej} it
was suggested to use mirror symmetry for Calabi-Yau fourfolds 
to study F-theory models with heterotic dual (see also \cite{Katz:1997eq}). In the following 
we will show how mirror symmetry can be applied to 
the geometries studied in this paper, which, however, do not admit 
a heterotic dual. 

Let us start by considering the mirror fourfold $\ov Y_{G}^*$ to the resolved space 
$\ov Y_{G}$. Our aim is to determine the gauge group obtained by compactifying 
F-theory on $\ov Y_{G}^*$. Hence we have to impose that $\ov Y_{G}^*$ itself is 
elliptically fibered. In fact, this is the case for the explicit elliptically fibered 
Calabi-Yau fourfolds used for GUT model-building  
given by two constraints \cite{Blumenhagen:2009yv,Grimm:2009yu}, 
as well as the elliptically fibered Calabi-Yau 
hypersurfaces studied in \cite{Klemm:1996ts}. This can be traced back to the 
fact that these Calabi-Yau spaces are realised in a toric ambient 
space. Thus their mirror \cite{Batyrev:1994hm,batyrev-1994-1,batyrev-1994-2,Kreuzer:2001fu} and fibration structure \cite{Avram:1996pj} can 
be analysed in detail using toric techniques (see ref.~\cite{Grimm:2009ef,Grimm:2009yu}, which includes 
a review of these techniques, and references). We first analyse the mirror of 
the space $\ov Y_{E_8}$ with a resolved $E_8$ singularity over $S$. 
The gauge group $\mathcal{H}(\ov Y_{E_8}^*)$ associated with the resolved elliptic fibration of $\ov Y_{E_8}^*$  
can be explicitly determined for the examples considered in this work, and will be of rather high rank. 
We can proceed in the 
same way for $\ov Y_G$, i.e.~the space in which the gauge group $E_8$ on 
$S$ is unfolded to $G$. One then shows that the new mirror gauge 
group is 
\beq \label{cH_factors}
  \mathcal{H}(\ov Y_{G}^*)  =  \mathcal{H}(\ov Y_{E_8}^*) \times \mathcal{H}\ .
\eeq
Here the new factor $\mathcal{H}$ is composed out of the 
structure groups \eqref{H-groups} of the bundle $V$ appearing over $S$ as well as the enhancement groups over 
curves and points:
\beq \label{gen_calH}
  \mathcal{H} = H^{k_S} \times H_C^{k_C} \times ... \times H^{k_P}_P \times ... \ .
\eeq
The dots indicate that one has to consider the bundle groups over 
all possible enhancement curves and points in $S$. Note that 
this picture makes direct contact with the spectral cover description and 
its extensions of section \ref{SCCs}. The precise form of $\mathcal{H}$ determines $P_V$
and vise versa. The largest group $H$ determines $G$ and hence the form of $P_V$ to be picked 
out of table \ref{spec_table}. The exponents in \eqref{gen_calH} are best explained 
by considering a specific example for $G,H$, as we will do next.

Given the $P_V$ in table \ref{spec_table} encoding the bundles on the divisor $S$
one can count the number of monomials in each of the defining $b_{n}$. Let us explain this 
for the example of $G = SU(5)$. Clearly, if $b_n=0$ for all  $n>0$ one obtains an $E_8$ gauge 
group or $SU(1)$ bundle. Let $k_n$ denote the number of possible non-zero monomials in $b_n$. Starting 
with an $SU(1)$ bundle, there are $k_2$ deformations to an $SU(2)$ bundle, $k_3$ deformations
to an $SU(3)$ bundle and $k_4$ deformations to an $SU(4)$ bundle. Finally, one has $k_6$ 
possible $SU(5)$ bundles corresponding to the different monomials in $b_6$.
One can then show that in the mirror fourfold $\ov Y_{SU(5)}^*$ one finds 
as gauge factor in \eqref{cH_factors} the group  
\beq \label{cH_SU(5)}
  \mathcal{H} = SU(5)^{k_6} \times SU(4)^{k_4} \times SU(3)^{k_3} \times SU(2)^{k_2} \times SU(1)^{k_0} \ .
\eeq
This gauge group can be determined by the Tate algorithm implemented via  
toric methods \cite{Candelas:1996su,Bershadsky:1996nh,Candelas:1997eh}.
Very basically, the weighted projective space 
$\mathbb{P}_{1,2,3}$ is encoded torically by the vertices
\beq
   \nu_1 = (0,-1),\quad \nu_2 = (-1,0), \quad \nu_3 = (3,2), 
\eeq
which correspond to the $x,y,z$ coordinates in the Tate equation \eqref{Tate1}. 
In the above construction the mirror manifold $\ov Y^*$ admits the dual two-torus
as the generic elliptic fiber. It given by the vertices  
\beq
\label{dualfiber1}
   \nu^*_1 = (1,-2),\quad \nu^*_2 = (-1,1), \quad \nu^*_3 = (1,1),
\eeq
which correspond to the $x,y,z$ coordinates in \textit{mirror} the Tate 
model. It is this dual Tate model in which one reads off the gauge 
group $\mathcal H$. 

Let us discuss this result in more general terms. Firstly, the exchange of $G$ 
and $\mathcal{H}$ under mirror symmetry arises as a simple combinatorial fact intrinsic
to elliptic fibrations with generic fiber $\mathbb{P}_{1,2,3}[6]$ of $E_8$ type.
In other words, it is possible to show that 
the identification \eqref{cH_factors} is rooted in the application of mirror 
symmetry for reflexive polyhedra and does not rely on the duality to a heterotic model. 
Hence, mirror symmetry will likely turn out to be a powerful tool to 
argue for the global validity of the spectral cover construction for 
Calabi-Yau examples in which all non-abelian gauge dynamics localises on $S$.
In particular, the split \eqref{chi-formula} of the Euler characteristic appears to 
be in accord with the factorisation of the dual gauge group  \eqref{cH_factors}.

\subsection{Split spectal covers and mirror symmetry for $U(1)$-restricted Tate models}
\label{GlobalSCCs}

In section \ref{fiber-restriction} we discussed an interesting specialisation of the 
Tate model by demanding that globally $a_6=0$. This led to 
a Calabi-Yau fourfold $X$ which admits an additional singularity 
over a curve $C$. We have argued that then an abelian factor $U(1)_X$
remains un-higgsed and can forbid dangerous dimension-4 proton decay operators. 
We now seek to apply mirror symmetry to the resolved manifold $\ov X$
and generalise the discussion of section \ref{mirror_spec}.
In particular, we want to determine the dual gauge group $\mathcal{H}$
for a singular fourfold $X_{SU(5)}$ with $SU(5)$ singularity over 
$S$ and $SU(2)$ singularity over $C$.

To begin with, we note that the restriction $a_6=0$ can be
implemented at the level of the Tate equation \eqref{Tate1}
by introducing a new coordinate $s$ with appropriate scaling 
relations to forbid a term $z^6 a_6$. This has been done in 
\eqref{Tate2_res}, where we also noted that the divisor $s=0$ corresponds
to the blow-up divisor $\hat D_C$ of the singular curve $C$. In 
contrast to a generic $\mathbb{P}_{1,2,3}[6]$ elliptic fiber, 
one now has an elliptic fiber encoded by the two-torus vertices
\beq \label{T2_2}
   \{\nu_i\} = \{ (0,-1),\, (-1,0),\, (3,2),\, (-1,-1)\}.
\eeq 
The new vertex $\nu_4=(-1,-1)$ corresponds to the coordinate $s$
and restricts the Tate form to be \eqref{Tate2_res}. 
Note that this fourfold $X$ still admits the section $z=0$ corresponding 
to the base $B$.

The analysis of the mirror $\ov X^*$ of the Calabi-Yau fourfold $\ov X$ 
proceeds as before.
 In particular, one can perform the restriction $a_6=0$
for the GUT examples of refs.~\cite{Blumenhagen:2009yv,Grimm:2009yu}  and show that the mirror is 
again elliptically fibered. The generic two-torus fiber of $\ov X^*$ 
is the dual to the fiber of $\ov X$ and can hence be inferred from \eqref{T2_2}
to be 
\beq \label{T2_2dual}
   \{\nu^*_i\} = \{ (1,-2),\, (-1,1),\, (1,0),\, (0,1)\}.
\eeq
By comparison with the dual of the $\mathbb{P}_{1,2,3}[6]$ 
elliptic fiber \eqref{dualfiber1} one notes that the vertex $(1,1)$ corresponding 
to the mirror $z$-coordinate has split into two vertices $(1,0),(0,1)$. 
We denote the corresponding coordinates by $z_1$ and $z_2$.
This implies that the corresponding Tate form is modified as
\bea \label{ref_Tate}
  P_{\rm T}^* &=&  P_1 + P_{U(1)}\\
  P_1 &=&  y^2 z_2 + x^3 z_1 +  a_{1,1}^* x y z_1 z_2 +  a^*_{2,1} x^2 z_1^2 z_2  
     + a_{2,2}^*y z_1^2 z_2^2 + a_{3,2}^* x z_1^3 z_2^2 + a^*_{4,3} z_1^4 z_2^3  \nonumber  \\
  P_{U(1)} &=&  a_{0,0} x^2 y\ .
\eea
This form of $P_{\rm T}^*$ is inferred by using the scaling relations of the 
coordinates $(x,y,z_1,z_2)$, which are the linear relations among the points \eqref{T2_2dual}.
Note that in contrast to the standard Tate model \eqref{Tate1}
based on $\mathbb{P}_{1,2,3}[6]$ one finds a term proportional to $x^2 y$. This is
precisely the deformation mirror dual to the blow-up $\hat D_C$. Hence, the 
extra term $P_{U(1)}$ signals the existence of the extra $U(1)$. In 
the restriction $a_6=0$ with no further modification $a_{0,0}$ is 
a single monomial and corresponds to a single $U(1)_X$ gauge boson. 

Since the elliptically fibered fourfold $\ov X^*$ has the form \eqref{ref_Tate} we 
have to reinvestigate the dual gauge group $\mathcal{H}$. 
Using the scaling relations to set $z_1 = 1 = z_2$, one finds that $P_1$ has the standard Tate form \eqref{Tate1}.
Hence, we can study the gauge group by analysing the vanishing of the polynomial 
over different divisors using the Tate algorithm. One shows that the dual gauge group naturally 
splits as 
\beq \label{cH_factors_2}
   \mathcal{H}(\ov X^*_{G}) =  \mathcal{H}(\ov X^*_{E_7}) \times \mathcal{H}_X\ .
\eeq
This is the analogue of \eqref{cH_factors} for the Tate model without the 
additional $U(1)$ factor. Note that 
now the maximal gauge group attained on $X_{G}$ is $E_7$ as stressed in sections \ref{Globalfac} and \ref{Bosons}. 
This matches nicely with the split \eqref{cH_factors_2} and can be checked for numerous examples. 
In particular, for $\ov X^*_{SU(5)}$ one finds 
\beq \label{cH_2}
    \mathcal{H}_X = SU(4)^{k_6} \times SU(3)^{k_4} \times SU(2)^{k_3} \times SU(1)^{k_2} \ .
\eeq 
Here $k_n$ are the number of monomials in the $b_n$ of the split spectral cover, 
which are equal to the $k_n$ in \eqref{cH_SU(5)}. 
Note that $k_0=0$  in agreement with the fact that one has set $a_6=b_0=0$.
In contrast to the dual gauge group in \eqref{cH_SU(5)} one thus finds that 
each factor has been broken by a $U(1)$. This is the global analogue of the 
split spectral cover construction $S[U(4) \times U(1)]$.

\section{Conclusions}

In this article we have analysed aspects of abelian gauge symmetries in global F-theory models. 
We have proposed a mechanism to guarantee un-higgsed $U(1)$ factors by a special restriction of 
the form of the Tate model. The presence of the $U(1)$ factor has been explained from different 
perspectives: The $U(1)$-restricted Tate model gives rise to localised massless states charged 
under the abelian group which signal the un-higgsing of the abelian symmetry. The abelian gauge 
boson can also be detected directly due an increase in $h^{1,1}$ of the Calabi-Yau fourfold after resolution. 
Finally in models with a IIB description the $U(1)$ can be traced back to the presence of a 
brane-image brane pair. We have been able to match one sector of the St\"uckelberg mechanism in 
Type IIB orientifolds with the appearance of the extended node in the affine Dynkin diagram in 
the fibers, a picture which we will describe in greater detail in the upcoming \cite{progress1}.

As a phenomenologically relevant application this mechanism allows one to implement a global 
$U(1)_X$ symmetry in F-theory GUT models that forbids dimension-4 proton decay. Crucially, our 
analysis goes beyond the split spectral cover approach, which is insensitive to the global 
question of $U(1)$ symmetries. The price one has to pay for the presence of $U(1)_X$ is a 
decrease of the Euler characteristic of the fourfold, which directly enters the D3-tadpole cancellation condition. 

An important open question concerns the precise definition of the gauge flux. Along the 
Coulomb phase of the underlying $E_8$, and, respectively, $E_7 \times U(1)$ Tate model 
one can easily study the Cartan fluxes, which are then transformed into non-abelian 
flux data after deformation to the actual Tate model of interest. We have argued that 
for $U(1)$ restricted Tate models a D-term naturally appears due to extra available 
flux data that involves a four-form written as the wedge product of two two-forms. 
The most pressing remaining question in this context is to identify the four-form 
describing the $U(1)$ flux concretely in terms of the resolved geometry in order to 
reliably compute its D3-tadpole charge and D-term. Progress is underway \cite{progress1}.

In the last part of the paper we have worked out the connection between the spectral 
cover construction and the global Tate model of an elliptic fourfold. We have argued 
that the  $E_8$ structure underlying a generic Tate model is responsible for the 
relevance of spectral covers, independently of a local gauge theory description 
or a heterotic dual.
This picture has been corroborated by an analysis of the mirror dual fourfolds, 
both for generic and for $U(1)$ restricted Tate models.
 Let us stress that  mirror symmetry is used here as a calculational tool
and appears not to correspond to a physical duality.
Clearly, this is in sharp contrast to heterotic/F-theory duality for which one 
expects a map of all physical quantities. It will be of considerable interest to 
focus on examples which admit no heterotic dual and highlight their
global properties.

\bigskip

\subsection*{Acknowledgements} 

We gratefully acknowledge discussions with R.~Blumenhagen, A. Collinucci, F.~Denef, A.~Hebecker, H.~Jockers, B.~Jurke, N.~Saulina, S.~Sch{\"a}fer-Nameki, A.~Klemm, D.~Klevers, 
S.~Krause, D.~Morrison and H.~P.~Nilles. 
We thank the Max-Planck-Institute in Munich for hospitality during parts of this work. TG also would like to thank the Harvard
theory group for hospitality. 
This research was supported in part by the SFB-Transregio 33 ``The Dark Universe'' by the DFG  and the National Science Foundation under Grant No. PHY05-51164.


\bigskip


\end{document}